\documentclass[12pt]{article}
\usepackage{cite}
\usepackage{graphicx}
\usepackage{multicol}
\usepackage{amsfonts}
\usepackage{amssymb}
\usepackage{amsmath}
\usepackage{heck}%
\usepackage{hyperref}
\numberwithin{equation}{section}

\def\bZ{\mathbb{Z}}
\def\bR{\mathbb{R}}
\def\bC{\mathbb{C}}
\newcommand{\cO}{\mathcal{O}}

\newcommand{\cN}{\mathcal{N}}

\DeclareMathOperator{\SU}{\mathit{SU}}

\newcommand{\rep}[1]{\mathbf{#1}}
\newcommand{\Tr}{\, {\rm Tr}}
\def\XX{\text{X}}
\def\IR{\text{IR}}
\def\UV{\text{UV}}
\def\GUT{\text{GUT}}
\def\singlet{\text{GUT}}

\begin{document}

\date{September 2010}
\title{ $\cN=1$ SCFTs from Brane Monodromy}


\authors{Jonathan J. Heckman$^1$,
Yuji Tachikawa$^1$, \\[4mm]
Cumrun Vafa$^2$,
and Brian Wecht$^1$

\bigskip\bigskip \normalsize

${}^{1}$School of Natural Sciences, Institute for Advanced Study, Princeton, NJ 08540, USA

\medskip

\texttt{jheckman,yujitach,bwecht@ias.edu}

\medskip

${}^{2}$Jefferson Physical Laboratory, Harvard University, Cambridge, MA 02138, USA

\medskip

\texttt{vafa@physics.harvard.edu}

}%

\abstract{We present evidence for a new class of
strongly coupled $\cN=1$ superconformal field theories (SCFTs)
motivated by F-theory GUT constructions. These SCFTs arise
from D3-brane probes of tilted seven-branes which undergo
monodromy. In the probe theory, this tilting corresponds to an $\mathcal{N} = 1$ deformation
of an $\mathcal{N} = 2$ SCFT by a matrix of
field-dependent masses with non-trivial branch cuts in the eigenvalues.
Though these eigenvalues characterize the geometry, we find that they do not
uniquely specify the holomorphic data of the physical theory.
We also comment
on some phenomenological aspects of how these theories
can couple to the visible sector. Our construction can be
applied to many $\cN=2$ SCFTs, resulting in a large new class of $\cN=1$ SCFTs.
}

\maketitle

\enlargethispage{\baselineskip}

\setcounter{tocdepth}{2}
\tableofcontents

\section{Introduction}

The interplay between string theory and geometry provides a
rich template for realizing many quantum field theories of
theoretical and potentially experimental interest. A common theme
is how geometric insights translate to non-trivial field theory statements, and
conversely, how statements about the field theory allow us to probe details of the
string geometry.

One class of theories which have recently been extensively studied is based on compactifications
of F-theory, in part because such constructions combine the flexibility of intersecting D-brane configurations
with the more attractive features of GUT models. See \cite{DWI,BHVI,WatariTATARHETF,BHVII,DWII}
and the references in \cite{Heckman:2010bq} for a partial list of work on F-theory GUTs.
 E-type geometric singularities play an especially
important role in realizing aspects of a GUT model such as the $\rep5 \times \rep{10} \times \rep{10}$ Yukawa coupling.
In most cases, the focus of such models has been to realize weakly-coupled field theories which reproduce
at least the qualitative features of the Standard Model. These models can also accommodate hidden sectors, which
can be added as separate sectors.

D3-branes provide an additional set of ingredients which are
present in such constructions. The presence of background fluxes often causes the D3-branes
to be attracted to the E-type points of the geometry \cite{FunParticles}. An important
feature of such points is that the axio-dilaton $\tau_\text{IIB}$ is of order one,
and thus the D3-brane worldvolume theory is strongly coupled.
Depending on the details of the geometry, we expect to realize a wide variety of possible strongly coupled quantum field theories.
The study of such field theories is of independent interest,
but is additionally exciting because the proximity to the visible sector suggests a phenomenologically novel way
to extend the Standard Model at higher energy scales.

D3-brane probes of F-theory singularities have been considered in various works, for example \cite{Banks:1996nj,Aharony:1996bi,Douglas:1996js,Fayyazuddin:1997cz,Fayyazuddin:1998fb,Aharony:1998xz,Noguchi:1999xq}. In many cases of interest, the probe theory becomes an interacting superconformal field theory (SCFT).
Geometrically, we engineer these SCFTs by considering the D3-brane probe of a parallel stack of seven-branes with gauge symmetry $G$.
When the stack is flat, this provides a geometric realization of rank 1 $\cN = 2$ SCFTs with flavor symmetry $G$, where $G$ can be $E_{6,7,8}$.
Denoting by $z_1,z_2$ the coordinates parallel to the stack, and by $z$ the coordinate transverse to the  stack,
in the probe theory $z$ becomes a chiral superfield $Z$ parameterizing the Coulomb branch, and $z_{1,2}$ become a decoupled hypermultiplet $Z_{1,2}$.
In addition, we have chiral operators $\cO$  in the adjoint representation of $G$ parameterizing the Higgs branch. When we have a
weakly coupled UV description of the theory, these operators can be written as composites made from quarks, e.~g.~$\cO \sim Q \widetilde Q$.

We study $\cN=1$ deformations of these theories by tilting the seven-branes.
This tilting is described by activating a position-dependent vev for an adjoint-valued scalar $\phi(z_1 , z_2)$ on the stack.
In the probe theory, this tilting corresponds to the superpotential
deformation \cite{FunParticles}:
\begin{equation}\label{deltaL}
\delta W =  \Tr_{G} (\phi(Z_1 , Z_2) \cdot \cO ).
\end{equation}
The eigenvalues of $\phi$ specify the location of the seven-branes.
Geometrically,
this tilting process is known as ``unfolding a singularity,'' and is specified purely in terms of the Casimirs of $\phi$.
A given matrix $\phi$ will satisfy a characteristic equation of the form:
\begin{equation}
\phi^{n} + b_{1}(z_{1} , z_{2}) \phi^{n-1} + \cdots + b_{n}(z_{1} , z_{2}) = 0
\end{equation}
where the $b_i(z_{1} , z_{2})$'s depend on the coordinates $z_{i}$.
The  most generic possibility is therefore that
the eigenvalues for $\phi$ will have branch cuts, a phenomenon
known as ``seven-brane monodromy''.  Such monodromies are a natural
ingredient for F-theory GUT models \cite{Hayashi:2009ge,BHSV,DWIII,EPOINT,Marsano:2009gv}.

Although the unfolding is dictated purely by the Casimirs of $\phi$,
we find that distinct $\phi$-deformations with the same Casimirs can produce strikingly different behavior in the IR.
In other words, the eigenvalues of $\phi$ are \emph{not} enough to specify the holomorphic data of the physical theory.
This freedom opens a new avenue for realizing intersecting seven-brane configurations, which to this point appear to have been relatively unexplored.\footnote{Some
discussion of the massless open string spectrum for ``exotic'' intersecting brane configurations based
on a nilpotent Higgs field has appeared in \cite{Donagi:2003hh}. Related issues in the context of F-theory compactifications will be discussed in \cite{CCHV}.}

In this paper, our aim will be to elucidate these differences from the point of
view of a probe D3-brane. Along the way, we will provide evidence for a large class of new
$\cN = 1$ deformations of $\cN = 2$ theories.\footnote{Though the setup is quite similar to that discussed in \cite{Aharony:1996bi}, here
we emphasize the eight-dimensional gauge theory interpretation of these deformations, some aspects of which cannot be seen from the Calabi-Yau fourfold alone.
Moreover, we find that the scaling dimensions of operators differ from what was found in \cite{Aharony:1996bi}, a point we shall comment more on in Appendix~\ref{subsec:FTT}.} We will provide various consistency checks that these deformations
lead to new interacting $\cN = 1$ SCFTs. For example, assuming we realize an
SCFT, we can use $a$-maximization \cite{Intriligator:2003jj} to determine
the infrared R-symmetry. We can also check that the scaling dimensions of operators
remain above the unitarity bound, and that the central charges of the SCFT
decrease monotonically after further deformations of the theory. Moreover, in some cases we can argue
that a further deformation induces a flow to a well-known interacting $\cN=2$ SCFT. In
such cases, the $\phi$-deformed theory can be viewed as an intermediate SCFT between the original $\cN = 2$ SCFT
and another IR $\cN=2$ SCFT.

The rest of the paper is organized as follows. In section \ref{sec:SETUP} we
introduce the brane setup for realizing the SCFTs of interest. As a first example, in section \ref{sec:D4probe} we discuss the deformation of $\cN=2$ $\SU(2)$ theory with four flavors, corresponding to a D3-brane probing a $D_4$ singularity.
Next, in section \ref{sec:ONEDEF} we turn to $\cN = 1$ deformations of a broader
class of non-Lagrangian theories, determining a general expression for the IR R-symmetry.
We also discuss some of the geometric content associated with this class of deformations.
Section \ref{sec:NILP}  considers specific examples of $\cN = 1$ deformations.
In section \ref{sec:POLY} we briefly consider further deformations of such theories
by  superpotential terms fixing the vev of the $Z_{i}$ and $Z$, and in section \ref{sec:SM} we indicate very briefly how the coupling of the SCFT sector to the visible sector works.
Section \ref{sec:CONC} contains our conclusions.
In Appendix~\ref{subsec:FTT} we briefly review some standard tools from field theory.

\section{Geometric preliminaries} \label{sec:SETUP}

In this section we review the general setup of D3-branes probing an F-theory singularity. Our aim here is to explain how the geometry of an F-theory compactification filters down to a D3-brane probe theory.

We are interested in F-theory singularities filling $\bR^{3,1}$.
The neighborhood of a singularity is a small patch in $\bC^3$ parameterized by $z$, $z_1$ and $z_2$.
Over each point  is an auxiliary elliptic curve, whose complex structure modulus is $\tau_\text{IIB}$. The elliptic curve is
given in Weierstrass form by:
\begin{equation}
y^2 = x^{3} + f(z_{1} , z_{2} , z) x + g(z_{1} , z_{2} , z),
\end{equation}
where the coefficients $f$ and $g$ fix $\tau_\text{IIB}$.

The locations of the seven-branes are specified by the zeros of the discriminant:
\begin{equation}
0 = \Delta(z_1,z_2,z) \equiv 4 f^{3} + 27 g^{2}.
\end{equation}
Each irreducible factor of $\Delta$ determines a hypersurface in $\bC^3$.
Throughout this paper we shall be interested in the behavior of a D3-brane probing a seven-brane located originally at $z = 0$.
In this convention, the coordinates $z_1$ and $z_2$ denote directions parallel to the seven-brane.

Away from the seven-branes, the worldvolume theory of a D3-brane is given by a $U(1)$
gauge theory, with holomorphic gauge coupling $\tau_{D3} = \tau_\text{IIB},$
whose value is controlled by the position of the D3-brane.
From the viewpoint of the open strings, the change in the coupling reflects the renormalization effects from the 3-7 strings.
As the D3-brane moves close to a seven-brane, some of these states will become light, and in some cases we expect to recover an interacting SCFT.

When the seven-brane is flat,  the probe theory is an $\cN = 2$ SCFT.
These theories are described
by a D3-brane probing a single parallel stack of seven-branes extending along $z_{1,2}$.
The  gauge symmetry on the seven-brane translates
to a global symmetry $G$ of the D3-brane probe theory.
The probe theory has a Coulomb branch parameterized by $z$,
the position of the D3-brane transverse to the seven-brane. The dimension of $z$ is given in Table~\ref{tableTIME}.
The Higgs branch of the probe theory corresponds to dissolving the D3-brane into the seven-brane stack as a gauge flux.
Examples of such probe theories include $\SU(2)$ with four flavors \cite{Banks:1996nj}, strongly-coupled theories
of the type of Argyres and Douglas \cite{Argyres:1995xn}, and the E-type theories found in  \cite{MNI,MNII}.

\begin{table}\[
\begin{array}
[c]{|c|c|c|c|c|c|c|c|}\hline
& H_{0} & H_{1} & H_{2} & D_{4} & E_{6} & E_{7} & E_{8}\\\hline\hline
\Delta & 6/5 & 4/3 & 3/2 & 2 & 3 & 4 & 6\\\hline
\end{array}\]
\caption{The scaling dimension of the Coulomb branch parameter for the $\mathcal{N} = 2$ SCFTs realized by a D3-brane
probing an F-theory singularity at constant dilaton. Here, the $H_{i}$ for $i=0,1,2$ respectively
correspond to the Argyres-Douglas theories arising from an $\SU(2)$ gauge theory with $i+1$ flavors. \label{tableTIME}}
\end{table}

Our main interest in this paper is to engineer $\cN = 1$ theories by considering more general  seven-brane configurations. The positions of the seven-branes
are dictated by a complex scalar $\phi$ on the seven-brane stack taking values in the adjoint representation of $G$.
Letting $\phi$ depend on $z_{1,2}$ corresponds to tilting the original stack of seven-branes.
Geometrically, this corresponds to performing a deformation of the original Weierstrass model via:
\begin{equation}
y^2 = x^3 + (f_{0} + \delta f(z_{1} , z_{2} , z)) x + (g_{0} + \delta g(z_{1} , z_{2}, z)).
\label{n1curve}
\end{equation}
Terms in $\delta f$ and $\delta g$ correspond naturally to expressions
made from the Casimirs of $G$ \cite{KatzMorrison}.
In the D3-brane probe theory this shows up as the superpotential deformation:
\begin{equation}
\label{origdef}
\delta W =  \Tr_{G} (\phi(Z_{1} , Z_{2}) \cdot \cO )
\end{equation}
where $\phi$ and $\cO $ are both  in the adjoint representation of $G$,
and $G$-invariant information like the positions of the seven-branes is characterized by the Casimirs of $\phi$. Thus, from a given $\phi$ one can naturally construct $\delta f$ and $\delta g$.

This deformation breaks $\cN = 2$ supersymmetry to $\cN=1$ when $[\phi , \phi^{\dag}] \neq 0$,
but is admissible as a background field configuration of the seven-brane once gauge fluxes are taken into account \cite{BHVI,FunParticles}.
Note that this deformation couples the original $\cN=2$ theory to the hypermultiplet $Z_{1,2}$.
The $\cN=1$ theory then has a moduli space parameterized by $z$ and $z_{1,2}$,
on a generic point of which the low energy limit is just a $U(1)$ theory.
The physical coupling of this low-energy $U(1)$ vector multiplet is holomorphic in $z$ and $z_{1,2}$, and is given
by a family of curves \cite{Intriligator:1994sm}. In this F-theoretic setup the required family of curves is exactly the elliptic fibration \eqref{n1curve}.
It is worth noting that in contrast to the $\cN=2$ case, this curve no longer
describes a full solution of the low-energy theory because the behavior of the chiral multiplets is no longer controlled by the gauge coupling.
The homogeneity of the curve can still be used to fix the relative scaling of further
mass deformations and the Coulomb branch parameters. We shall meet examples of this analysis later on.

Although $\phi$ should be holomorphic without branch cuts, its eigenvalues can have branch cuts as we vary $z_{i}$. In cases where such
structures exist, we say these deformations exhibit ``seven-brane monodromy.''
A simple example is the matrix:
\begin{equation}\label{Z2mono}
\phi=\left[
\begin{array}
[c]{cc}%
0 & 1\\
Z_{1} & 0
\end{array}
\right]
\end{equation}
which has eigenvalues $\pm\sqrt{Z_{1}}$.
Closely related to  seven-brane monodromy is the generic
presence of nilpotent mass deformations.
For example, in equation (\ref{Z2mono}), the constant contribution is a mass matrix which is upper triangular, and thus nilpotent.
This is the source of the monodromy  after a further deformation by a lower triangular part proportional to $Z_1$.

Such nilpotent mass deformations are already of independent interest. Indeed, because
all of the Casimirs of a nilpotent matrix vanish, the $\cN = 1$  curve of the deformed theory is identical to the original $\cN = 2$ curve.
But this apparent invariance of the holomorphic geometry is deceptive.
Clearly, we have added a mass term to the theory which breaks $\cN = 2$ supersymmetry
in the probe theory and moreover gives a mass to some of the degrees of freedom of the original theory.
Thus, we see that the Calabi-Yau fourfold alone does \emph{not} fully specify the holomorphic data of the compactification.
This does not immediately contradict the standard lore in much of the F-theory
literature that the Calabi-Yau fourfold and flux data are enough to specify
the compactification. The point is that Casimirs of $\phi$ and of the gauge flux are not enough.
This is quite exciting from the perspective of F-theory compactifications, because it
points to a far greater degree of flexibility in the specification of a
compactification, based on more holomorphic data than just the Casimirs of $\phi$.

The greater freedom in specifying $\phi$ is also connected with the presence of singular fibers in the
Calabi-Yau geometry. Indeed, in a compactification in which all singularities of the geometry
have been deformed away, our general expectation is that there is no ambiguity in reconstructing
a unique choice of $\phi$. From the perspective
of the seven-brane gauge theory this is equivalent to asking whether a given
characteristic equation for $\phi$ uniquely determines the physics of the seven-brane configuration. It would be interesting to study whether this natural
physical expectation is always met for a general unfolding.

Before going further, let us point out that there is no distinction among $z$, $z_1$, and $z_2$ from the ten-dimensional point of view.
For example, consider the configuration
\begin{equation}
y^2 = x^3 + z^5 + z_1.
\end{equation}
This can be thought of as either a deformation of an $E_8$ seven-brane at $z=0$,
or as a deformation of an $H_0$ seven-brane at $z_1=0$.
This suggests that the same $\cN=1$ theory can be realized by deformations of two different $\cN=2$ SCFTs. We hope to come back to this question in the future.

\section{Probing a $D_{4}$ Singularity}\label{sec:D4probe}

The cases of E-type flavor symmetry in which we are interested do not have an
obvious Lagrangian description. As a warm-up, we start in this section by
studying $\cN=1$ deformations of a D3-brane probing a $D_{4}$ singularity of F-theory; this setup leads to a Lagrangian theory.
The weakly coupled theory of a D3-brane probing a $D_{4}$ singularity is given by an
$SU(2)$ gauge theory with four quark flavors $Q_{i} \oplus \widetilde{Q}_{\overline i}$
for $i=1,...,4$ \cite{Banks:1996nj}.
The superpotential is dictated by $\cN=2$ supersymmetry:%
\begin{equation}
W=\sqrt{2}Q_{i}\varphi\widetilde{Q}_{\overline{i}}.
\end{equation}
The theory has an $SO(8)$ flavor symmetry.

The moduli space is characterized in terms of gauge-invariant operators built
from the elementary fields. The Coulomb branch of the theory is parameterized
by the coordinate:
\begin{equation}
Z=\frac{1}{2}\Tr_{SU(2)}\varphi^{2}.
\end{equation}
Next consider the Higgs branch, touching the Coulomb branch at $Z = 0$.
The Higgs branch is parameterized in terms of composite meson operators  $\cO$ quadratic in the quarks.
They transform in the adjoint representation of $SO(8)$.
It is convenient to decompose them in terms of  irreducible representations of
$U(4)\subset SO(8)$:
\begin{align}
\rep{16}  &  :\cO _{i\overline{j}}=Q_{i}\widetilde{Q}_{\overline{j}} &
\rep{6}  &  :\cO _{[ij]}=Q_{[i}Q_{j]}&
\overline{\rep{6}}  &  :\cO _{[\overline{ij}]}=\widetilde{Q}_{[\overline{i}}%
\widetilde{Q}_{\overline{j}]}.
\end{align}
Although at this stage we can write the $\cO $'s in terms of the $Q$'s, when we later explore
E-type theories, we will have nothing to work with except the analogous $\cO $ operators.

In addition to the degrees of freedom described above, there is a free hypermultiplet $Z_1\oplus Z_2$ representing the position of the D3-brane
parallel to the seven-brane.
Thus, we initially  have two decoupled CFTs. Tilting the seven-branes to some new
configuration couples these two CFTs, and generates a non-trivial flow to a
new $\cN=1$ theory.

In F-theory, the geometry of a $D_{4}$ singularity
is given by the Weierstrass equation:%
\begin{equation}
y^{2}=x^{3}+Az^{3}+xz^{2} \label{SWD4}%
\end{equation}
where $A$ is a free parameter. The modulus $\tau$ of the torus \eqref{SWD4}, depending on $A$, gives the coupling of the $\SU(2)$ theory.

In the remainder of this section we study in greater detail $\cN = 1$ deformations
of the $D_{4}$ probe theory. In particular, our aim will be to present evidence that these theories
realize interacting SCFTs in the IR. For the most part, we focus on nilpotent mass deformations such that $\phi$ takes
values in a single Jordan block of $SU(n) \subset U(4) \subset SO(8)$. Many of the checks we perform in the following subsections
can be viewed as elucidating more details of the interacting SCFT.

\subsection{Mass Deformations and the $\cN=1$ Curve}\label{sec:homo}

Mass deformations of the theory correspond to deformations of the form $\Tr_{SO(8)} (\phi \cdot \cO )$
for $\phi$ independent of $Z_1$ and $Z_2$. The Casimirs of $\phi$ determine
deformations of the original $\cN=2$ curve:%
\begin{equation}
y^{2}=x^{3}+Az^{3}+xz^{2}+\left(  f_{2}z+f_{4}\right)  x+g_{4}z+g_{6}\label{N1curve}
\end{equation}
where the $f_{i}$'s and $g_{i}$'s correspond to degree $i$ polynomials in the
masses $m$ formed from expressions built from
the Casimirs of $\phi$. Using the formulation in terms of the $\cO $'s,
these deformations can be written as:%
\begin{equation}
\delta W=m_{j\overline{i}}\cO _{i\overline{j}}+m_{[\overline{ij}]}\cO _{[ij]}%
+m_{[ij]}\cO _{[\overline{ij}]}.
\end{equation}

Additionally, we can consider field-dependent mass deformations which couple
the $\cN=2$ $D_{4}$ theory to the free hypermultiplet $Z_1\oplus Z_2$.
From the perspective of the geometry, the only change is that now the $f_{i}$
and $g_{i}$ in \eqref{N1curve} can depend on the coordinates $z_{1}$ and $z_{2}$.

Let us now consider a deformation by the nilpotent mass term:
\begin{equation}
\delta W=m_{1\overline{2}}\cO _{2\overline{1}}=m_{1\overline{2}}Q_{2}\widetilde {Q}_{\overline{1}}.
\end{equation}
As the mass terms are nilpotent matrices, all Casimirs built from these operators are
trivial, and the $\cN=1$ curve is identical to the $\cN=2$ curve.\footnote{
In the $\cN=1$ theory, we have added mass terms for some of the quark flavors.
As the theory flows from the UV to the new IR theory, the beta function of the gauge coupling is
non-zero. This raises the question: Since we have initiated a flow of the gauge coupling,
why does the $\cN = 1$ curve predict that on the Coulomb branch of the
deformed theory there is no change to the value of $\tau$?

To see what is happening, let us note that there are three mass
scales of interest. First, there is the mass scale $m$ associated with the
nilpotent deformation. In addition, there is the scale $m_{W}$ of the W-bosons of
the Higgsed gauge theory. At scales $m_{W}<\mu<m$, we have integrated out a quark
flavor from the $SU(2)$ gauge theory, and the coupling increases as the theory
flows to the IR. However, below the scale $\mu<m_{W}$, the W-bosons of the $SU(2)$
gauge theory are also massive, and this in turn counters the effects of the
initial decrease. In particular, we see that as the theory flows to the scale
$\mu=m_{W}^{2}/m<m_{W}<m$, the value of the gauge coupling has flowed back to its
original value. All of this behavior is automatically encoded in the geometry. See
\cite{Argyres:1999xu} for related discussions.}

The existence of an $\cN=1$ curve  constrains the relative scaling dimensions
of operators in the deformed theory, as in the $\cN=2$ case.
Since the $\cN=1$ curve is no different from the $\cN=2$ curve,
this implies that the relative scalings of $z$ and the
mass deformations in the new $\cN = 1$ theory obey the same relations
as in the original $\cN = 2$ theory.
Let us now check how this works from the viewpoint of the Lagrangian.

To this end, consider the effective superpotential:
\begin{equation}\label{Potpot}
W_{eff}=\sum_{i=3,4}\sqrt{2}Q_{i}\varphi\widetilde
{Q}_{\overline{i}}-2\frac{Q_{1}\varphi\varphi\widetilde{Q}_{\overline{2}}%
}{m_{1\overline{2}}}
\end{equation}
after integrating out $Q_2$ and $\widetilde Q_{\overline 1}$.

Assuming we have flowed to an interacting CFT, let us now compute the relative
scaling dimensions of the various mass deformations. For simplicity, we
restrict attention to the mass parameters transforming in the adjoint
representation of $U(4)$. Taking the superpotential of equation (\ref{Potpot})
to be marginal in the IR, we learn that the dimensions of the $Q$'s are related to the dimension
of $z$ by:
\begin{align}
[Q_{1}\widetilde{Q}_{\overline{2}}] &= 3-\Delta_\IR, &
[Q_{1}\widetilde{Q}_{\overline{I}} ]=[Q_{I}\widetilde{Q}_{\overline{2}}] &= 3-\frac{3}{4}\Delta_\IR  , &
[Q_{I}\widetilde{Q}_{\overline{J}}] &= 3-\frac{1}{2}\Delta_\IR
\end{align}
where $\Delta_\IR  $ is the scaling dimension of $z$ and $I,J = 3,4$.
The corresponding mass terms are
\begin{equation}
\delta W=m_{2\overline{1}}Q_{1}\widetilde{Q}_{\overline{2}}+m_{I\overline{1}%
}Q_{1}\widetilde{Q}_{\overline{I}}+m_{2\overline{I}}Q_{I}\widetilde
{Q}_{\overline{2}}+m_{J\overline{I}}Q_{I}\widetilde{Q}_{\overline{J}}%
\end{equation}
where the dimension of the $m$'s can be easily obtained from the data given above.
In order to match these mass deformations to quantities of the $\cN=1$
curve, we must form invariants under the surviving flavor symmetries.
We obtain four invariants with corresponding dimensions:
\begin{align}
[m_{2\overline{1}}] = [m_{I\overline{J}} m_{J\overline{I}}] &= \Delta_\IR, &
[m_{2\overline{I}} m_{I\overline{1}}] &= \frac{3}{2}\Delta_\IR , &
[m_{2\overline{I}}m_{I\overline{J}}m_{J\overline{1}}] &=2\Delta_\IR.
\end{align}

We now compare these invariants to invariants of the $\cN = 2$ theory.
First note that these expressions transform non-trivially under the (now broken) original flavor symmetry.
Including appropriate factors of $m_{1 \overline{2}}$ to form flavor invariants of the original
$\cN = 2$ theory, we see that these expressions descend from the $\cN = 2$
theory Casimirs:
\begin{align}
[m_{1\overline{2}}m_{2\overline{1}}] = [m_{I\overline{J}}m_{J\overline{I}}] &= \Delta^{\cN=2} , &
[m_{1 \overline{2}} m_{2\overline{I}} m_{I\overline{1}}] &= \frac{3}{2}\Delta^{\cN=2}, \\
[m_{1\overline{2}} m_{2\overline{I}}m_{I\overline{J}}m_{J\overline{1}}] &=2\Delta^{\cN=2}.
\end{align}
Note that the same relative scalings are obtained once we set $m_{1\overline
{2}} =1$, as appropriate upon treating $m_{1 \overline{2}} Q_{2} \widetilde{Q}_{\overline{1}}$ as a marginal
operator in the IR theory. Similar considerations hold for other group theory invariants, and for more general
deformations as well.

\subsection{$a$-Maximization and Nilpotent Mass Deformations\label{AmaxNilp}}

Nilpotent mass deformations are of particular interest because although they constitute a non-trivial
$\cN=1$ deformation of the theory, they do not alter the geometry of
the $\cN=1$ curve.
For simplicity, we again confine our analysis to nilpotent deformations where $\phi$
takes values in $SU(n)\subset U(4)\subset SO(8)$. Explicitly, we
consider the mass deformations:%
\begin{equation}
\delta W=\sum_{k=1}^{n-1} m_{k\overline{k+1}%
}\cO _{k+1\overline{k}}.%
\end{equation}

Let us first consider in detail the case $n=2$. Upon integrating out the
quarks $Q_{2}\oplus\widetilde{Q}_{\overline{1}}$, we are left with an $SU(2)$
gauge theory with chiral superfields $\varphi$, $Q_{3}\oplus\widetilde{Q}_{3}%
$, $Q_{4}\oplus\widetilde{Q}_{4}$, and $Q_{1}\oplus\widetilde{Q}_{\overline{2}
}$ with an effective superpotential \eqref{Potpot}. By inspection, one finds
\begin{equation}
R(Q_3)=R(\widetilde{Q}_{\overline{3}})=
R(Q_4)=R(\widetilde{Q}_{\overline{4}}),\qquad
R(Q_{1}) =R(\widetilde{Q}_{\overline{2}}).
\end{equation}
We require that the IR R-symmetry is non-anomalous,
and that the two superpotential terms in equation (\ref{Potpot})
are marginal in the IR.
We find that $R(Q_3)$ and $R(Q_1)$ can be expressed in terms of
$R(\varphi)$, which can then be determined by  $a$-maximization.
The calculation is straightforward;  we find a local maximum
at $R(\varphi)=(-9+\sqrt{145})/6\simeq0.51$. As can be checked,
all gauge-invariant operators have dimensions above the unitarity bound.

Let us now generalize this to the cases $n=3,4$. Upon integrating
out the heavy quarks, we are left with the effective superpotential:
\begin{align}
W_{eff}^{(n=3)}  &  =\sqrt{2}Q_{4}\varphi\widetilde{Q}_{\overline{4}}%
+2\sqrt{2}\frac{Q_{1}\varphi^{3}\widetilde{Q}_{\overline{3}}}{m_{1\overline
{2}}m_{2\overline{3}}}, &
W_{eff}^{(n=4)}  &  =-4\frac{Q_{1}\varphi^{4}\widetilde{Q}_{\overline{4}}%
}{m_{1\overline{2}}m_{2\overline{3}}m_{3\overline{4}}}.
\end{align}
Again we impose the conditions that  $R_\IR$ is non-anomalous  and  that the
effective superpotential is marginal in the IR.  For both $n=3,4$ we find
one undetermined parameter which we fix by $a$-maximization.

The behavior of the $n=2,3$ theories is quite similar.
The case of $n=4$ presents a new
phenomenon, in that here it would appear that there are unitarity bound violations.
Indeed, assigning R-charges to the operators $Z$ and
$Q_{1} \widetilde{Q}_{\overline{4}}$ then requires either that both operators saturate the unitarity bound, or that one
operator violates this bound.

What are we to make of this case?  One possibility
is that this theory may not be an interacting conformal theory.
This does not appear very plausible, because nothing
drastic appears to be happening to the geometry. For example, we can still move onto the Coulomb branch and compute a non-trivial
dependence of $\tau$ on the parameters $z_{i}$ and $z$. In what follows, we shall assume
that much as in \cite{Seiberg:1994pq}, an accidental symmetry appears which
rescues only this individual operator.

We  now recompute the dimensions for the
quarks in the IR theory under the assumption that only $Z$ decouples
in the IR, with an associated emergent $U(1)$ which only acts on $Z$. This emergent $U(1)$ can be
included in $a$-maximization via the procedure described in \cite{Kutasov:2003iy}.
The scaling dimensions for the various
fields are then given by the values shown in Table~\ref{table0}.
\begin{table}\[
\begin{array}
[c]{|c|c|c|c|c|c|c|c|c|c|c|}\hline
& R(\varphi) & [Z] & [Q_{1}] & [\widetilde{Q}_{\overline{4}}] &
[\widetilde{Q}_{\overline{3}}] & [\widetilde{Q}_{\overline{2}}] &
[\widetilde{Q}_{\overline{1}}] & [Q_{4}] & [Q_{3}] & [Q_{2}]\\\hline\hline
\cN=2 & 2/3 & 2 & 1 & 1 & 1 & 1 & 1 & 1 & 1 &
1\\\hline
n=2 & 0.51 & 1.52 & 0.74 & 1.12 & 1.12 & 0.74 & \XX & 1.12 &
1.12 & \XX\\\hline
n=3 & 0.36 & 1.07 & 0.69 & 1.23 & 0.69 & \XX & \XX & 1.23 & \XX &
\XX\\\hline
n=4 & 0.25 & 1 & 0.76 & 0.76 & \XX & \XX & \XX & \XX & \XX & \XX\\\hline
\end{array}\]
\caption{Dimensions of the elementary fields obtained from nilpotent deformations of the probe $D_{4}$ theory.
The operators of the CFT are specified by gauge invariant expressions built from these elementary fields. Entries with an ``X'' indicate
fields which have been integrated out of the low energy theory.\label{table0}}
\end{table}
Note that although the dimensions of the $Q$'s are less than one, all of the composite
operators built from two $Q$'s have dimension above the unitarity bound.

\subsection{Large $N$ Limit}

In the previous section we discussed the specific case of a single D3-brane
probing a $D_{4}$ singularity. In the context of brane constructions, it is
natural to consider the theory obtained by $N$ D3-branes probing the same
configuration.
The theory is given by an  $\cN=2$ $USp(2N)$ gauge theory with four quark flavors $Q_i \oplus \widetilde Q_{\overline i}$ and an additional hypermultiplet $P \oplus \widetilde P$ in the two-index antisymmetric representation \cite{Douglas:1996js,Fayyazuddin:1998fb}.
The $\cN=2$ superpotential for this theory is:%
\begin{equation}
W=\underset{i=1}{\overset{4}{\sum}}\sqrt{2}Q_{i}\varphi\widetilde
{Q}_{\overline{i}}+\sqrt{2}P\varphi\widetilde{P}.
\end{equation}
This is an $\cN=2$ SCFT.
In addition, there is a free hypermultiplet $Z_{1}\oplus Z_{2}$ describing the motion of the center of mass for the configuration.

Let us now deform this theory by nilpotent masses. Assuming that the IR limit is an $\cN=1$ SCFT without any accidental symmetry, we
perform $a$-maximization and expand the result to first order in $1/N$.
The R-charge assignment for $\varphi$ and the scaling dimensions
for the elementary fields are given in Table~\ref{tableN}.
\begin{table}\[
\begin{array}
[c]{|c|c|c|c|c|}\hline
& R(\phi) & [Z] & [P]=[Q] & [Q^{\prime}]\\\hline\hline
\cN=2 & \frac{2}{3} & 2 & 1 & 1\\\hline
n=2 & \frac{2}{3}-\frac{2}{3N} & 2-\frac{2}{N} & 1+\frac{1}{2N} &
\frac{1}{2}+\frac{1}{N}\\\hline
n=3 & \frac{2}{3}-\frac{8}{3N} & 2-\frac{8}{N} & 1+\frac{2}{N} &
\frac{6}{N}\\\hline
n=4 & \frac{2}{3}-\frac{20}{3N} & 2-\frac{20}{N} & 1+\frac{5}{N} &
-\frac{1}{2}+\frac{20}{N}\\\hline
\end{array}\]
\caption{Dimensions of the elementary fields for the large $N$ limit of the probe $D_4$ theories with a nilpotent mass
deformation turned on. Note that although the dimensions of some elementary fields fall far below the
unitarity bound, the scaling dimensions of composite operators involving these fields can still remain above this bound.\label{tableN}}
\end{table}

In this table, $Q$ indicates any flavors that couple to $\varphi$ through $Q \varphi \widetilde Q$, and
$Q'$ indicates flavors that couple through $Q' \varphi^n \widetilde Q'$, which as in the previous section
is present after integrating out the massive quarks.
Looking at the table, we see that for $n>2$, some of the operators quadratic in
$Q'$ will fall below the unitarity bound. As before, we can assume that these
operators become free fields and decouple from the IR\ theory and re-do $a$-maximization.
However, because in the large $N$ limit the number of offending operators is $\cO(N^0)$, there is no change at this order to the R-charge assignments, since $a$ is $\cO(N^2)$.

We still believe that this system flows to an IR SCFT.
Consider the related deformation by $\phi^{\prime} = \phi + \phi^{T}$,
given by symmetrizing the original deformation. In the UV $\cN = 2$ theory, this corresponds to a deformation which preserves
$\cN = 2$ supersymmetry, and leads to the theory describing $N$ D3-branes probing $H_{0,1,2}$ singularities.
If we now consider deforming first by $\phi$, and after some long RG time deforming further by $\phi^{T}$, we then flow
to this same $\cN = 2$ theory. Assuming that this further deformation can be added at an arbitrarily late RG time,
this strongly suggests that the above description of the $\cN = 1$ theory is legitimate, and that the further
deformation by $\phi^{T}$ induces a flow from the $\phi$-deformed theory to the $\phi^{\prime} = \phi + \phi^{T}$ deformed theory.
The only concern here would be that somehow the further deformation
by $\phi^{T}$ is not valid in the deep IR, though this seems rather implausible.
Furthermore, nothing dramatic seems to be
happening to the F-theory geometry, so it seems reasonable to assume that
the CFT is still present.

\section{$\cN=1$\ Deformations: Generalities\label{sec:ONEDEF}}

In the previous section we studied the theory of a D3-brane probing a $D_{4}$
singularity, relying on a Lagrangian formulation of the theory to analyze the
effects of various deformations. In many cases of interest, however, we do not
have a weakly coupled Lagrangian description, as when  we have a flavor symmetry $G = E_{n}$.

In this section we consider superpotential deformations of the form:
\begin{equation}
\delta W=\text{ Tr}_{G}(\phi(Z_{1},Z_{2})\cdot \cO ). \label{deform}%
\end{equation}
Our main assumption will be that we obtain a CFT in the
infrared, and we shall present some consistency checks of this statement.
Assuming we do flow to a new CFT, it is important to determine the scaling
dimensions of operators, and the values of the various central charges in the
infrared. This is of intrinsic interest, but is also of interest in potential
model building applications where the degrees of freedom from the D3-brane
couple to visible sector degrees of freedom associated with modes localized on seven-branes.

In the case where $\phi$ has no constant terms,
the results of \cite{Green:2010da} establish that this deformation
is marginally irrelevant, and so induces a flow back to the original CFT \cite{FunParticles}.
For this reason, we shall focus on the case where $\phi$ has a non-trivial constant part. Further, since
we are interested in the structure of deformations where the geometric singularity is retained
at $z_i = 0$, we also demand that all Casimirs of $\phi$ vanish at $z_i = 0$. Hence,
the constant part of $\phi$ is a nilpotent matrix. This reinforces the point that nilpotent deformations
go hand in hand with generic deformations of an F-theory singularity.

We now describe the general procedure for obtaining the R-charge assignments
for the matter fields after turning on a combination of relevant and marginal
deformations.
We first catalogue the symmetries of the UV theory, and then the
surviving symmetries compatible with the deformation $\Tr_{G}(\phi \cdot \cO )$.
The $SU(2) \times U(1)$ R-symmetry of the $\cN=2$ theory contains two $U(1)$'s, which we
call $R_{UV}$ and $J_{\cN=2}$; see Appendix~\ref{subsec:FTT} for details.
The remaining global symmetries are the non-abelian flavor symmetry $G$ and the $U(1)$ generators
which rotate the fields $Z_{i}$ of the free hypermultiplet. Since
the infrared R-symmetry is a linear combination of abelian symmetries, it is
enough to focus on the Cartan subalgebra $U(1)^r $ of the flavor symmetries $G$, where $r$ is the rank of $G$.
We shall denote by $F_{i}$ the corresponding generators, where
$i$ runs from $1$ to $r$.
Finally, we denote by $U_{i}$ the generator under
which $Z_{i}$ has charge $+1$. Under $R_\UV $, $J_{\cN = 2}$ and the $U_{i}$, the
charges of the original operators are given in Table~\ref{somers}.
\begin{table}\[
\begin{array}
[c]{|c|c|c|c|}\hline
& \cO  & Z & Z_{j}\\\hline\hline
R_\UV  & 4/3 & \left(  2/3\right)  \Delta_\UV (Z) & 2/3\\\hline
J_{\cN=2} & -2 & 2\Delta_\UV (Z) & -1\\\hline
U_{i} & 0 & 0 &  \delta_{ij} \\\hline
\end{array}
\]
\caption{Assignment of UV charges where $\Delta_\UV (Z)$ denotes the
dimension of $Z$ in the UV. The charges refer to the bottom component of
each supermultiplet. \label{somers}}
\end{table}

The $\cN=1$ deformation explicitly breaks some of these flavor
symmetries. The infrared R-symmetry will then be given by a linear combination
of the UV symmetries, and possibly some additional emergent flavor symmetries.
In what follows, we shall assume that there are no
emergent abelian symmetries in the infrared. Then the R-symmetry is given by
\begin{equation}
R_\IR=R_\UV +\left(  \frac{t}{2}-\frac{1}{3}\right)  J_{\cN%
=2}+\underset{i=1}{\overset{r}{\sum}}t_{i}\cdot F_{i}+u_{1}U_{1}+u_{2}%
U_{2}.
\end{equation}
The coefficient of $J_{\cN=2}$ has been chosen for later
convenience.

Let us determine which symmetries are left unbroken by the original deformation. At
first, it may appear that no solution is available which is compatible with a
deformation of the form given by equation (\ref{deform}). Indeed, though there
are at most $r+ 3$ flavor symmetries, there will typically be far more
independent entries in the matrix $\phi$. However, in the infrared, our
expectation is that some of these deformations will become irrelevant
operators. For example, if one of the entries of $\phi$ contains a term of the
form $Z_{1}^{100}$, we expect this term to be irrelevant in the infrared.
This also matches with geometric expectations. The geometry
is well-approximated to leading order  by the lowest degree polynomials in the
$Z_{i}$. Higher order polynomials correspond to subleading features of the
geometry. Since we are only interested in a small neighborhood of the region
where $Z_{1}=Z_{2}=0$, much of this information is washed out
in the infrared. We shall return to this theme later when we discuss the
UV and IR behavior of the characteristic polynomial for $\phi$.

The flow to a new CFT is dominated by the operators of lowest scaling
dimension. In the UV, the most relevant terms are the constant matrices. By assumption,
the constant matrix is nilpotent, and so by a unitary change of basis, we can present
it as an upper triangular matrix. For simplicity, in what follows we assume that the constant part
of $\phi$ denoted by $\phi_{0}$ decomposes as a collection of
$n_{a} \times n_{a}$ blocks, each of which corresponds to an upper triangular matrix:
\begin{equation}\label{DECOMP}
\phi_{0} = \underset{a=1}{\overset{k}{\oplus}}J^{(a)}.
\end{equation}
We assume that the upper triangular matrices $J^{(a)}$ are generic in the sense that the first superdiagonal
has only nonzero entries.

Associated with each block is an $SU(2)$ subalgebra of the original flavor symmetry group $G$
with generators $T_{\pm}^{(a)}$ and $T_{3}^{(a)}$ in the spin
$j_{(a)}=(n_{(a)}-1)/2$ representation, satisfying
\begin{align}
\lbrack T_{+}^{(a)},T_{-}^{(a)}]  &  =2T_{3}^{(a)} &
\lbrack T_{3}^{(a)},T_{\pm}^{(a)}]  &  =\pm T_{\pm}^{(a)}.
\end{align}
In this basis, the $T_{3}^{(a)}$ generator is:%
\begin{equation}
T_{3}^{(a)} = \text{diag}(j_{(a)},j_{(a)}-1,...,1-j_{(a)},-j_{(a)}).\label{Tj}
\end{equation}

Most of the data associated with each block $\phi^{(a)}_{0}$ of the decomposition in equation \eqref{DECOMP} drops
out in the infrared. To see how this comes about, consider the entries of the upper triangular block $J^{(a)}$.
Along each superdiagonal of the matrix, the value of the $T_{3}^{(a)}$ charge is the same.
Moving out from the diagonal, the entries of $\phi^{(a)}_{0}$ on the first
superdiagonal have charge $+1$, the second have $+2$, and so on until the upper
righthand entry which has charge $n_{(a)} - 1$. The operators $\mathcal{O}^{(a)}$
which pair with $\phi^{(a)}_{0}$ in the deformation have respective $T^{(a)}_{3}$ charges $-1$
down to $-(n_{(a)} - 1)$. Since all operators on the same superdiagonal have the same $T_{(3)}^{(a)}$ charge,
we see that the one with the charge of smallest norm will dominate the flow. In other words, the first superdiagonal
of $\phi^{(a)}_{0}$ dominates the flow. In the following we assume $\phi^{(a)}_{0}$ takes the
form of a nilpotent Jordan block from the start.

Then the operators $\cO ^{(a)}=\Tr_{G} (\phi_{0}^{(a)} \cdot \cO )$ have $T_3^{(a)}$
charge $-1$.
The requirement that $\cO ^{(a)}$ is marginal for all $a$ in the IR can now be satisfied by the choice
\begin{equation}
\label{riris}
R_\IR=R_\UV +\left(  \frac{t}{2}-\frac{1}{3}\right)  J_{\cN%
=2}-t T_{3}+u_{1}U_{1}+u_{2}U_{2}%
\end{equation} where \begin{equation}
T_{3}=\sum_{a} T_3^{(a)}
\end{equation} is the generator of the diagonal $\SU(2)$ subalgebra.
The coefficients $u_{i}$ are still undetermined.
We can now organize the operators $\mathcal{O}$ into representations of this diagonal $SU(2)$. We denote by
$\mathcal{O}_{s}$ an operator with spin $s$ under this $SU(2)$.

To fix the value of the $u_{i}$'s, we need to know which of the remaining operator
deformations are most relevant in the IR. In the case of a
deformation by a constant $\phi$, the free hypermultiplet
decouples, and we can neglect the $U_{i}$'s.
We therefore focus on the additional
effects of $Z_{i}$-dependent deformations.
Unitarity dictates that  $\cO $ and the $Z_{i}$ have dimensions greater than or equal to one.  This means that if two or more $Z_{i}$'s multiply an operator $\cO $, their product
will be irrelevant. Hence, it is enough to focus on contributions which are linear in the $Z_{i}$.

Most of the deformations linear in the $Z_{i}$ will also be irrelevant.
Given two operators $Z_{i}\times \cO _{s}$ and $Z_{i}\times \cO ^{\prime}_{{s}^{\prime}}$
which have different $T_{3}$ charges $s$ and $s^{\prime}$, the operator with the larger charge
will have lower dimension, and will therefore dominate the flow.

Since the IR behavior is dictated by the operators $\cO _{s}$ with the highest values of $s$, it is
enough to consider the deformation by just these highest values. Let $\cO_{S_{1}}$ and $\cO_{S_{2}}$ denote the operators  which respectively multiply $Z_{1}$ and $Z_{2}$.
The parameters $u_{i}$ are now fixed by requiring that these deformations have R-charge 2 in the IR. In terms of $S_{1}$, $S_{2}$ and $t$, this constraint yields:
\begin{equation}
u_{i} = \left(S_{i} + 3/2 \right) t - 1 \equiv \mu_{i} t - 1 \qquad\text{where}\quad \mu_i=S_i+3/2.\label{mui}
\end{equation}
We now see that for a given $\mu_1, \mu_2$, the only free parameter in $R_\IR$ is $t$.

\subsection{The $\cN = 1$ Curve and Relative Scaling Dimensions}\label{sec:CURVE}

Now let us study to what extent we can read off properties of the $\cN = 1$ deformed
theory without determining $t$.
As we have already mentioned,
on the Coulomb branch of the $\cN = 1$ theory,
we can read off the $U(1)$ coupling from the $\cN=1$ curve,
which is the F-theory geometry \eqref{n1curve}.
Homogeneity of this $\cN = 1$ curve then predicts the relative scaling dimensions of the mass deformations to that of the Coulomb branch parameter.

The form of the infrared R-symmetry \eqref{riris} implies
\begin{equation}
\Delta_\IR(Z)    =\frac{3}{2}t\times\Delta_\UV (Z).
\end{equation}
Therefore the unknown parameter $t$ can be eliminated in favor of the ratio $\rho=\Delta_\IR(Z)/\Delta_\UV(Z)$, and we find
\begin{align}
\Delta_\IR(Z_{i})  &  =\left(  S_{i} - \frac{1}{2}\right)\rho, &
\Delta_\IR(\cO _{s})  &  =3-(s+1)\rho. \label{Os}
\end{align}
The dimension of the mass parameter $m_{\cO _{s}}$ associated with an operator $\cO _{s}$ is then:
\begin{equation}
\Delta_\IR( m_{\cO _{s}}) =3-\Delta_\IR(\cO _{s})=(s+1)\rho.
\end{equation}
In particular, when we form flavor invariants out of the mass parameters $m$
as in section \ref{sec:homo}, their ratio in the IR is the same in the UV,
because the total spin $s$ of the flavor invariants is zero.

As a passing comment, let us also
note that the value of the IR central charge $k_\IR$
agrees with the computation in \cite{FunParticles}: using \eqref{k}, we easily find
\begin{equation}
k_\IR=\rho k_\UV.
\end{equation}

\subsection{Characteristic Polynomials in the Infrared} \label{ssec:IRchar}

Some aspects of the deformation $\phi$ are irrelevant in
the IR. To study the possibilities, we can consider choices
for $\phi$ which in the IR induce a flow to the same theory as the original
$\phi$. To indicate the UV and IR behavior we write $\phi_\UV $ and
$\phi_\IR$.

The matrix $\phi_\IR$ is fully characterized by terms which
are at most linear in the $Z_{i}$. Indeed, as the D3-brane
only probes a small patch of the geometry, it is insensitive to higher order
terms in the geometry, which are effectively gone in the deep infrared. Of course
these further effects can still be probed by moving a finite distance onto the Coulomb branch.

Given two different $\phi$'s, linearizing in the $Z_{i}$ can produce the same
IR behavior for $\phi$. For example, the characteristic equations
\begin{align}
\phi^{5}+z_{1}  &  =0,\\
\phi^{5}+z_{2}^{w}\phi + z_{1}  &  =0
\end{align}
respectively define solvable and unsolvable quintics. For $w>1$, however,
the term linear in $\phi$ drops out in the infrared. Thus,
the UV and IR behavior of $\phi$ can be different.

For the more mathematically inclined reader, we note that the
``seven-brane monodromy group'' corresponds to the Galois group for the characteristic equation for $\phi$.
The monodromy group acts by permuting the roots of the polynomial, and is
indicated by the specific branch cut structure present in the eigenvalues of
$\phi$. Here we see that the infrared monodromy groups which can be realized are
of quite limited type.

A polynomial of the form:%
\begin{equation}
\phi^{n}+b_{2}\phi^{n-2}+...+b_{n}=0 \label{charequS5}%
\end{equation}
will generically have maximal Galois group given by $S_{n}$, the
symmetric group on $n$ letters. In particular, we can take the $b_{i}$ to admit a
power series expansion in the $Z_{i}$. By a general coordinate redefinition of
the geometry, and a field redefinition in the CFT, we see that generically, we
can take the leading order behavior of the lowest coefficients to be
$b_{n}=Z_{1}$ and $b_{n-1}=Z_{2}$.

Finding a representative $\phi$ with the corresponding characteristic equation
is also straightforward. To illustrate the main points, let us focus on the
case of $\phi$ given by a $5\times5$ matrix. A representative $\phi$ with
characteristic equation as in (\ref{charequS5}) can be taken in the form:%
\begin{equation}
\phi=\left[
\begin{array}
[c]{ccccc}%
0 & 1 &  &  & \\
-c_{2}^{(2,1)} & 0 & 1 &  & \\
-c_{3}^{(3,1)} & -c_{2}^{(3,2)} & 0 & 1 & \\
-c_{4}^{(4,1)} & -c_{3}^{(4,2)} & -c_{2}^{(4,3)} & 0 & 1\\
-c_{5}^{(5,1)} & -c_{4}^{(5,2)} & -c_{3}^{(5,3)} & -c_{2}^{(5,4)} & 0
\end{array}
\right]
\end{equation}
for appropriate $c_{n}^{(i,j)}$. In the
infrared, the relevant deformation by $\phi$ is:
\begin{equation}
\phi_\IR=\left[
\begin{array}
[c]{ccccc}
& 1 &  &  & \\
&  & 1 &  & \\
&  &  & 1 & \\
-\alpha Z_{2} &  &  &  & 1\\
-Z_{1} & -\beta Z_{2} &  &  &
\end{array}
\right]
\end{equation}
for some coefficients $\alpha$ and $\beta$. The characteristic equation for
$\phi_\IR$ is:%
\begin{equation}
\phi_\IR^{5}+(\alpha +\beta)Z_{2}\phi_\IR+Z_{1}=0.
\end{equation}
As can be checked, the monodromy group for this degree five polynomial is
again $S_{5}$.


\subsection{Central Charges and $a$-Maximization \label{CENTRAL}}

We now fix the infrared R-symmetry using $a$-maximization.
The trial central charge $a_\IR(t)$ can be computed
using 't Hooft anomaly matching between the UV and IR theories.
Thus $a_\IR(t)$ depends on $a_\UV $, $c_\UV $ and $k_\UV $,
as well as the details of the Jordan block structure associated with the deformation $\Tr_{G}(\phi \cdot \cO )$.

Plugging \eqref{riris} into \eqref{boo} and rewriting it using \eqref{AAA}--\eqref{CCC},
we obtain the value of the IR central charges as follows:
\begin{align}
a_\IR  &  =\frac{3}{32}\left[
\begin{array}
[c]{c}%
\left(  36a_\UV -27c_\UV -\frac{9k_\UV r}{4}\right)  t^{3}+\left(
-72a_\UV +36c_\UV +\frac{9}{4}(u_{1}+u_{2})\right)  t^{2}\\
+\left(  48a_\UV -12c_\UV -\frac{9}{2}\left(  u_{1}^{2}+u_{2}^{2}\right)
\right)  t+\left(  -(u_{1}+u_{2})+3(u_{1}^{3}+u_{2}^{3})\right)
\end{array}
\right] \label{AIR}\\
c_\IR  &  =\frac{1}{32}\left[
\begin{array}
[c]{c}%
\left(  108a_\UV -81c_\UV -\frac{27k_\UV r}{4}\right)  t^{3}+\left(
-216a_\UV +108c_\UV +\frac{27}{4}(u_{1}+u_{2})\right)  t^{2}\\
+\left(  96a_\UV +12c_\UV -\frac{27}{2}\left(  u_{1}^{2}+u_{2}^{2}\right)
\right)  t+\left(  -5(u_{1}+u_{2})+9(u_{1}^{3}+u_{2}^{3})\right)
\end{array}
\right] \\
k_\IR  &  =\frac{3}{2}t\times k_\UV  \label{kir}%
\end{align}
where in the above, we have introduced the parameter $r$ which measures the sizes of the nilpotent block:
\begin{equation}\label{rdef}
r\equiv  2 \Tr\left(  T_{3}T_{3}\right).
\end{equation}

We need to find the local maximum of  $a_\IR$ given in (\ref{AIR}) to find the value $t$.
There are two cases of interest, which we analyze separately. The first
case corresponds to deformations where $\phi$ is a constant nilpotent matrix. In this case, we formally
set $u_1 = u_2 = 0$. In addition, we must remember that the free hypermultiplet $Z_{1} \oplus Z_{2}$
decouples, and in particular does not contribute to the central charges $a_\UV $ and $c_\UV $.
The other case corresponds to the more generic geometry in which $\phi$ has some
linear dependence in both $Z_1$ and $Z_2$. In this case, the contribution
from the hypermultiplet must be included in the values
of $a_\UV $ and $c_\UV $. These values are tabulated in Appendix~\ref{subsec:FTT}.

\paragraph{Nilpotent Mass Case}

First consider the case where $\phi$ is a constant nilpotent matrix. Setting
$u_{1}=u_{2}=0$ in \eqref{AIR},
$a$-maximization yields an extremum at:
\begin{equation}
t_{\ast}=\frac{4}{3}\times\frac{8a_\UV -4c_\UV -\sqrt{4c_\UV ^{2}%
+(4a_\UV -c_\UV )k_\UV r}}{16a_\UV -12c_\UV -k_\UV r}. \label{tstarnilp}%
\end{equation}
with $r$ as in equation (\ref{rdef}). Note that for $r=0$, we recover
$t_{\ast}=2/3$, corresponding to the correct branch of solutions to the
quadratic equation.

\paragraph{Monodromic Case}

Next consider the case of position-dependent $\phi(Z_{1},Z_{2})$ where a term linear in each $Z_{i}$ appears in the deformation $\Tr_{G}(\phi \cdot \cO )$.
Applying \eqref{mui} in \eqref{AIR} and performing $a$-maximization, we find
\begin{equation}
\label{tstar}
t_{\ast} = \frac{-B - \sqrt{B^{2}  - 4 A C}}{2 A}
\end{equation}
where:
\begin{align}
A &  = \frac{3}{4}(48a_\UV -36c_\UV -3k_\UV r+3\mu_{1}+3\mu_{2}-6\mu_{1}^{2}-6\mu_{2}%
^{2}+4\mu_{1}^{3}+4\mu_{2}^{3})  \\
B &  = -3  - 48a_\UV  + 24c_\UV  + 6\mu_{1} + 6\mu_{2} - 6\mu_{1}^{2} - 6\mu_{2}^{2} \\
C &  = -3 + 16a_\UV  - 4c_\UV +\frac{8}{3}\mu_{1}+\frac{8}{3}\mu_{2}.
\end{align}
The choice of branch cut in equation (\ref{tstar}) is fixed as in the nilpotent case.

\section{Probing an $E_n$ Singularity}\label{sec:NILP}

Having given a general analysis of the
expected IR R-symmetry, we now turn to some examples.
In fact the $D_4$ case analyzed in section \ref{sec:D4probe} falls
within the analysis presented in the last section.
Here we will study the $E_n$ case where a weakly-coupled UV description is not available.

We first consider nilpotent mass deformations of the $\cN=2$ $E_{8}$
SCFT.  We find a consistent structure of flows between various deformations of this theory. We also study the large $N$ limit of such probe theories.
After this analysis, we turn to the more generic case of deformations which include
a $Z_{i}$-dependent contribution. In F-theory, allowing a position-dependent profile
for the field $\phi$ corresponds to tilting the configuration of the seven-branes.
Finally, we consider some particular examples which are of interest for F-theory GUTs.

\subsection{Nilpotent Mass Deformations\label{E8NILP}}

We now turn to deformations of the $E_{8}$ theory by $\phi$ valued in
$SU(n)\subset SU(9) \subset E_{8}$.\footnote{This decomposition makes the representation
content under $G_\GUT $ less manifest, but for our present purposes this is
not necessary. 
} The adjoint of $E_{8}$ decomposes under $SU(9)$ into
the adjoint representation, and a three index antisymmetric tensor as:%
\begin{align}
\rep{248}  &  \rightarrow \rep{80}+\rep{84}+\overline{\rep{84}}.
\end{align}
Hence, the operators $\cO $ initially transforming in the
adjoint representation of $E_{8}$ will decompose into singlets, and one-, two-,
and three-index tensor representations of $SU(n)$. Another feature of interest
is that this also suggests a natural split between the cases of $n\leq5$ and
$n>5$. For $n\leq5$, the three index representation is already the dual
representation of a representation with a smaller number of indices, while for
$n>5$, no such redundancy is present.

For simplicity we confine our analysis to deformations where $\phi$ is given
by a single $n\times n$ Jordan block.
The parameter $r$ introduced in \eqref{rdef} is
then given by
\begin{equation}%
r= (n^{3}-n)/6.
\end{equation}

Let us now comment on the representation content of the operators $\cO $. Under the $SU(2)$ subalgebra specified by the
Jordan block, the fundamental representation becomes a spin $j=(n-1)/2$ irreducible representation of $SU(2)$.
For the higher tensor index structures,  the indices are free to range
over the spin $j$ irreducible representation, subject to appropriate anti-symmetry or tracelessness
conditions.
Since the dimension of the operators $\cO $ is specified by its spin content, and
thus its tensor structure in $SU(n)$, we shall denote by $\cO _\singlet $ the
singlets, $\cO _{i}$ an operator in the fundamental of $SU(n)$, $\cO _{ij}$ an
operator in the two-index antisymmetric, and so on. We denote by $\cO _{i}%
^{\min}$ the operator in the fundamental with the lowest scaling dimension,
with similar notation for the other $\cO $'s.
The scaling dimension of $\cO ^{\min}$ is then given by \eqref{Os} for appropriate $s$.

Using equation (\ref{tstarnilp}) and the expressions for the operator scaling
dimensions and the values of the central charges obtained in
section \ref{CENTRAL}, we find the values for the various parameters given in Table~\ref{table1}.
\begin{table}\[
\begin{array}
[c]{|c|c|c|c|c|c|c|c|c|c|c|c|}\hline
n & E_{8} & 2 & 3 & E_{7} & 4 & E_{6} & 5 & 6 & 7 & 8 &
9\\\hline\hline
t_{\ast} & \XX & 0.54 & 0.40 & \XX & 0.29 & \XX & 0.23 & 0.18 & 0.15 & 0.12 & 0.10\\\hline
a_\IR & 3.96 & 3.42 & 2.69 & 2.46 & 2.09 & 1.71 & 1.66 & 1.34 & 1.11 & 0.94 & 0.81\\\hline
c_\IR & 5.17 & 4.40 & 3.40 & 3.17 & 2.62 & 2.17 & 2.07 & 1.67 & 1.38 & 1.16 & 1.00\\\hline
k_\IR & 12 & 9.73 & 7.14 & 8 & 5.31 & 6 & 4.09 & 3.25 & 2.66 & 2.22 & 1.88\\\hline
\left[  Z\right]   & 6 & 4.86 & 3.57 & 4 & 2.65 & 3 & 2.04 & 1.63 & 1.33 & 1.11 & 1\\\hline
\lbrack \cO _\singlet ] & 2 & 2.19 & 2.41 & 2 & 2.56 & 2 & 2.66 & 2.73 & 2.78 & 2.81 & 2.84\\\hline
\left[  \cO _{i\overline{j}}^{\min}\right]   & \XX & 1.38 & 1.22 & \XX & 1.23 & \XX & 1.30 & 1.37 & 1.45 & 1.52 & 1.59\\\hline
\lbrack \cO _{i}^{\min}] & \XX & 1.78 & 1.81 & \XX & 1.89 & \XX & 1.98 & 2.05 & 2.11 & 2.17 & 2.22\\\hline
\lbrack \cO _{ij}^{\min}] & \XX & \XX & \XX & \XX & 1.67 & \XX & 1.64 & 1.65 & 1.67 & 1.70 & 1.75\\\hline
\left[  \cO _{ijk}^{\min}\right]   & \XX & \XX & \XX & \XX & \XX & \XX & \XX & 1.51 & 1.45 & 1.43 & 1.43\\\hline
\end{array}
\]
\caption{Central charges and operator scaling dimensions for the $\mathcal{N} = 1$ SCFTs realized by
nilpotent $\phi$-deformations of the $\mathcal{N} = 2$ $E_8$ SCFT. An ``X'' indicates that
this entry has no meaning for the specified deformation.\label{table1}}
\end{table}
In the table, we have ordered the
entries according to decreasing values of $a_\IR$. We have
also included the corresponding $\mathcal{N}=2$ SCFT values. Note that
increasing $n$ always decreases $a_\IR$, in accord with the expectation that
we lose degrees of freedom as we continue to flow to the IR. Further, the
theories with $\mathcal{N}=2$ supersymmetry and $E$-type flavor symmetry have
smaller central charges than their nilpotent counterparts with the same
non-abelian flavor symmetries ($n=2$ for $E_{7}$ and $n=3$ for $E_{6}$).
Physically this is reasonable, as we have given a mass deformation to a
smaller number of 3-7 strings in the case of nilpotent deformations. Finally,
in all cases but the last with $n=9$, all of the original operators remain
above the unitarity bound. In this one case, we find that a first application
of $a$-maximization yields a value for the dimension of $Z$ which falls below
the unitarity bound. In the above, we have assumed that there is an emergent
$U(1)$ which only acts on $Z$, so that $Z$ decouples as a free field.
Recomputing the value of the parameter $t$ and the associated dimensions
yields the corresponding values for $n=9$. Note that this behavior is quite
similar to what we observed in the case of the $D_{4}$ probe theory. In that
case, we observed that small nilpotent deformations produced a self-consistent
picture for operator dimensions, while for the  $4\times4$ Jordan
block, there was an apparent violation of the dimension for the operator $Z$.
There, this was ascribed to integrating out so many quarks, so presumably a
similar phenomenon is present in the E-type case as well.

\begin{figure}
[t!]
\begin{center}
\includegraphics[
height=2.8in
]%
{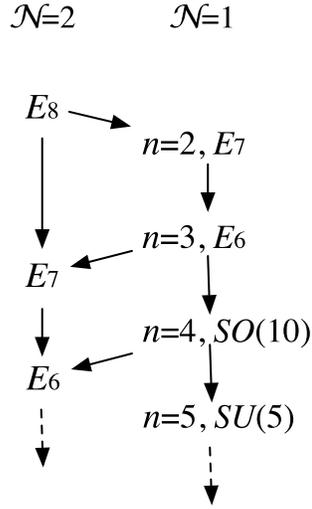}%
\caption{Depiction of the various theories obtained by performing a single
$n \times n$ nilpotent Jordan block deformation of the $E_{8}$ $\mathcal{N}=2$ SCFT.
Also indicated is the associated non-abelian flavor symmetry of each $\mathcal{N} = 1$ theory. As
summarized in table \ref{table1}, increasing the size of the block leads to a further
flow, and a decrease in the central charge $a_\IR$. In the $n=3$ theory the
non-abelian flavor symmetry is $E_{6}$, but it is nevertheless possible to
deform the theory to the $\mathcal{N}=2$ $E_{7}$ SCFT. Similarly, the $n=4$
theory with non-abelian flavor symmetry $SO(10)$ can be deformed to the
$\mathcal{N}=2$ $E_{6}$ SCFT.}%
\label{RG}%
\end{center}
\end{figure}

The theories defined by different $n$'s are all connected by further
deformations. Mathematically, starting from the deformation defined by an
$n \times n$ nilpotent Jordan block, there is a deformation we
can perform by enlarging $\phi$ to an $(n + 1) \times (n + 1)$ nilpotent
Jordan block. This corresponds to a further deformation by a relevant operator. For example,
starting from the $n = 2$ theory, adding the operator $O^{\rm{min}}_{i}$ corresponds
to adding the next entry of the $3 \times 3$ nilpotent block,
inducing a deformation to the $n = 3$ theory.
Note that from table \ref{table1} each such operator is relevant in the corresponding theory, so it will indeed induce a flow to a new theory. See figure \ref{RG} for a depiction of these flows.

Further deformations of the $\phi$-deformed theories can also induce flows back to an $\mathcal{N} = 2$ theory.
For example, the $n = 2$ theory is specified by deforming the $E_{8}$ $\mathcal{N} = 2$ theory by an operator $\mathcal{O}_{-}$
dotted into the matrix:
\begin{equation}
\phi_{0}=\left[
\begin{array}
[c]{cc}%
0 & 1 \\
0 & 0
\end{array}
\right]  .
\end{equation}
In the IR theory we can consider further deformations with opposite $T_3$ charge which we denote by $\cO_{+}$. In the UV theory, this would
correspond to adding the deformation:%
\begin{equation}
\phi=\phi_{0}+\phi_{0}^{T}=\left[
\begin{array}
[c]{ccc}%
0 & 1 \\
1 & 0
\end{array}
\right]  .
\end{equation}
This deformation preserves $\mathcal{N}=2$ supersymmetry since
$\left[  \phi,\phi^{\dag}\right] = 0$, and induces a flow to the $E_{7}$ $\mathcal{N} = 2$ SCFT.
This is consistent with the fact that the $n = 2$ theory has a larger central charge. Note also that the non-abelian
E-type flavor symmetries agree.

It is also possible to perform a further deformation of the nilpotent deformed
theories to an $\mathcal{N}=2$ theory with a \textit{larger} non-abelian flavor
symmetry. For example, a similar argument to that given for the $n = 2$ theory establishes
that in the $n = 3$ theory, we can perform a flow to an $\mathcal{N} = 2$ theory associated
with the $\phi$-deformation:
\begin{equation}
\phi=\phi_{0}+\phi_{0}^{T}=\left[
\begin{array}
[c]{ccc}%
0 & 1 & 0\\
1 & 0 & 1\\
0 & 1 & 0
\end{array}
\right]  .
\end{equation}
This induces a flow back to the $E_{7}$ theory which has a \textit{larger}
$E$-type flavor symmetry than the $n =3$ theory! Indeed, by a unitary change of basis we can write $\phi\simeq$ diag$(+\sqrt{2},-\sqrt{2},0)$, which has
commutant $E_{7}$. Note that this is also consistent with the fact that the
central charge of the $n=3$ theory is greater than that of the $\mathcal{N}=2$
$E_{7}$ theory.

Similar considerations hold for the $n=4$ theory, and deformations to the
$E_{6}$ theory. Indeed, starting from
\begin{equation}
\phi_{0}=\left[
\begin{array}
[c]{cccc}%
0 & 1 & 0 & 0\\
0 & 0 & 1 & 0\\
0 & 0 & 0 & 1\\
0 & 0 & 0 & 0
\end{array}
\right]  ,
\end{equation}
a further deformation by a lower triangular matrix will have rank three or
higher.\footnote{To see this, note that in the further deformation by such a
$\phi$, the columns of this matrix provide three linearly independent vectors
in the image space of $\phi$.} Thus, the $E$-type non-abelian flavor symmetry
is at most $E_{6}$. This is also consistent with the fact that the central
charge of the $n=4$ theory is lower than the $\mathcal{N}=2$ $E_{8}$ and
$E_{7}$ theories, but is higher than the $E_{6}$ theory; the $E_{6}$ theory can indeed be realized
by an appropriate deformation of the $n =4$ theory. Finally, let us note that for the $n = 5$ theory, similar
considerations establish that adding a further deformation to $\phi_{0}$ can allow the non-abelian flavor symmetry to
increase its rank by at most one unit, so that it is at most $SO(10)$. Note that this is
consistent with the fact that the central charge of this theory is \textit{below} that
of the $\mathcal{N} = 2$ $E_{6}$ theory. All of these checks suggest a highly non-trivial
structure, providing further evidence for the existence of the $\phi$-deformed
theories. See figure \ref{RG} for a schematic presentation of how these theories
are connected by further deformations.

\subsection{Large $N$ Limit}

It is quite natural to also consider the limit with a large number of D3-branes
located at the same point. Note that if we move all the D3-branes together away from the seven-branes, the
probe theory is $\cN=4$ $U(N)$ gauge theory, so we expect an
interacting SCFT at the origin of the Coulomb branch.

An interesting feature of the large $N$ case
is that for $n \geq 3$, we find that some of the operators $\cO $ will naively violate the unitarity
bound.
To see the apparent violations of the unitarity bound, we can compute the value of $t_{\ast}$
in the large $N$ limit, to find:
\begin{equation}
t_{\ast}=\frac{2}{3}-\frac{2\alpha}{3}\frac{1}{N}+\cO \left(
1/N^{2}\right)
\end{equation}
where $\alpha$ is an order one parameter which depends on the details of the
model. The coefficient $\alpha$ is positive, because the dimension of the
Coulomb branch parameter decreases along the flow. Indeed, the IR\ dimension
of $Z$ is:%
\begin{equation}
\Delta_\IR(Z)=\frac{3}{2}t\times\Delta_\UV (Z)=\Delta_\UV (Z)-\frac{\alpha
}{N}+\mathcal{\cO }\left(  1/N^{2}\right)  .
\end{equation}
Further, the dimension of the operators $\cO _{s}$ are:%
\begin{equation}
\Delta_\IR(\cO _{s})= 3-(s + 1)+\frac{\alpha}{N}\times
(s + 1)+\mathcal{\cO }\left(  1/N\right)  \text{.}%
\end{equation}
Thus, we see that although $Z$ essentially maintains its UV value, the
operators $\cO _{s}$ will generically fall below the unitarity bound for sufficiently
high values of $s$. Since $\alpha>0$, the operators which violate the
unitarity bound satisfy:%
\begin{equation}
s > 1.
\end{equation}
As the size of $n$ increases, the available spins $s$ will also increase,
and the number of operators falling below the unitarity bound will increase. For example,
starting from the $E_{8}$ $\cN=2$ SCFT, consider adding a deformation
by the $3\times3$ Jordan block $\phi_{0}$:%
\begin{equation}
\phi_{0}=\left[
\begin{array}
[c]{ccc}
& 1 & \\
&  & 1\\
&  &
\end{array}
\right]  .\label{zot}
\end{equation}
In this theory,  there is an operator with spin
$s = +2$ corresponding to the mass parameter in the lower left corner of \eqref{zot}.
The dimension of this operator is order $1/N$, and in particular,
below the unitarity bound.

Note, however, that we expect this deformation to induce a flow to a CFT.
Indeed, in the theory deformed by just the operators
$\cO _{-}$ of $T_{3}$ charge $-1$, the operators $\cO _{+}$ of $T_{3}$ charge $+1$
have dimension:
\begin{equation}
\lbrack \cO _{+}]=1+\frac{\alpha}{N}%
\end{equation}
which is slightly above the unitarity bound if we trust our naive calculation.
It seems therefore consistent to perform a further deformation by this operator, leading to a
flow to the large $N$ $\mathcal{N} = 2$ $E_{7}$ theory.
Performing $a$-maximization, we find to leading order in the $1/N$ expansion:
\begin{equation}
a_{\cN=2, E8}= \frac32N^2+\frac52N
>a_{\cN=1, E8,\phi}= \frac32N^2 - \frac12N
\gg a_{\cN=2, E7}= N^2+\frac32N.
\end{equation}

How then do we interpret the fact that the operator with $SU(2)$ spin $+2$
drops below the unitarity bound? A self-consistent possibility is that as
the dimension of the offending
operator decreases and passes to the unitarity bound, an additional $U(1)$
emerges, and the offending operator decouples as a free field, with dimension
$1$. Re-performing $a$-maximization with this extra $U(1)$ included, we can then read off
the new IR R-symmetry. Note, however, that since only $O(1)$ operators fall below the unitarity bound,
this is a second order effect in a $1/N$ expansion. Working to first order in a $1/N$ expansion, we
can therefore ignore this effect.
Finally, because the operators $\mathcal{O}^{\rm{min}}_{i}$ generically
decouple as free fields in the IR, it also follows that as opposed to the case $N = 1$,
the natural extension of flows from an $n \times n$ nilpotent Jordan block to
an $(n+1) \times (n+1)$ nilpotent Jordan block deformation is now obstructed.
Nevertheless, we find that increasing $n$ always decreases the value of the central charge $a_\IR$. Further, we find that all $\phi$-deformed $E_{8}$ theories have central charge above that
of the $\mathcal{N} = 2$ $E_{7}$ SCFT.

\subsection{Maximal Monodromy}

As our first example of non-trivial seven-brane monodromy, we consider a
$\phi$ taking values in $SU(n) \subset SU(9)\subset E_8$. Moreover, in this section we assume that the unfolding is ``generic'' in the
sense that the characteristic polynomial for $\phi$ has
Galois group $S_{n}$.\footnote{Let us note
that in the specific context of F-theory GUTs where
$\phi$ takes values in the $SU(5)_{\bot}$ factor of
$SU(5)_\GUT  \times SU(5)_{\bot} \subset E_{8}$, this choice is unacceptable, because it means
there is one curve for all fields in the $\overline{\rep{5}}$ and $\rep5$ of
$SU(5)_\GUT $, and in particular no distinction between the Higgs and matter
fields.}

Using the general results of section \ref{sec:ONEDEF}, and the values of the
UV central charges (including the contribution from the free hypermultiplet)
we can extract the infrared values of the central charges and infrared scaling
dimensions. To illustrate the general pattern, we now present a general table of
the IR\ values for $n=2,...,9$ for $\phi_\IR$ satisfying the characteristic
equation:%
\begin{equation}
\phi_\IR^{n}+z_{2}\phi_\IR+z_{1}=0
\end{equation}
for $n>2$. The UV\ inputs are quite similar to the case of the nilpotent
deformations, though we also need to specify the values of the parameters
$\mu_{1}$ and $\mu_{2}$. In the case of a single Jordan block, we have:%
\begin{align}
\mu_{1}  &  =(n-1)+\frac{3}{2}, &
\mu_{2}  &  =(n-2)+\frac{3}{2}.%
\end{align}
The IR\ values of the various dimensions and central charges are presented in Table~\ref{table2}.
\begin{table}\[
\begin{array}
[c]{|c|c|c|c|c|c|c|c|c|c|c|c|}\hline
n & E_{8} & 2 & 3 & E_{7} & 4 & E_{6} & 5 & 6 & 7 & 8 & 9\\\hline\hline
t_{\ast} & \XX & 0.53 & 0.39 & \XX & 0.29 & \XX & 0.22 & 0.17 & 0.14 & 0.12 & 0.10\\\hline
a_\IR & 3.96 & 3.41 & 2.69 & 2.46 & 2.09 & 1.71 & 1.66 & 1.35 & 1.13 & 0.96 & 0.83\\\hline
c_\IR & 5.17 & 4.38 & 3.40 & 3.17 & 2.61 & 2.17 & 2.05 & 1.67 & 1.39 & 1.19 & 1.03\\\hline
k_\IR & 12 & 9.60 & 7 & 8 & 5.15 & 6 & 3.93 & 3.10 & 2.52 & 2.10 & 1.76\\\hline
\left[  Z\right]   & 6 & 4.80 & 3.5 & 4 & 2.58 & 3 & 1.96 & 1.55 & 1.26 & 1.05 & 1\\\hline
\left[  Z_{1}\right]   & 1 & 1.60 & 1.75 & 1 & 1.72 & 1 & 1.64 & 1.55 & 1.47 & 1.40 & 1.32\\\hline
\left[  Z_{2}\right]   & 1 & 1 & 1.17 & 1 & 1.29 & 1 & 1.31 & 1.29 & 1.26 & 1.22 & 1.17\\\hline
\lbrack \cO _\GUT ] & 2 & 2.20 & 2.42 & 2 & 2.57 & 2 & 2.67 & 2.74 & 2.79 & 2.83 & 2.85\\\hline
\left[  \cO _{i\overline{j}}^{\min}\right]   & \XX & 1.40 & 1.25 & \XX & 1.28 & \XX & 1.36 & 1.45 & 1.53 & 1.60 & 1.68\\\hline
\lbrack \cO _{i}^{\min}] & \XX & 1.80 & 1.83 & \XX & 1.93 & \XX & 2.02 & 2.10 & 2.16 & 2.21 & 2.23\\\hline
\lbrack \cO _{ij}^{\min}] & \XX & \XX & \XX & \XX & 1.71 & \XX & 1.69 & 1.71 & 1.74 & 1.78 & 1.83\\\hline
\left[  \cO _{ijk}^{\min}\right]   & \XX & \XX & \XX & \XX & \XX & \XX & \XX & 1.58 & 1.53 & 1.51 & 1.53\\\hline
\end{array}\]
\caption{Central charges and operator scaling dimensions for the $\mathcal{N} = 1$ SCFTs realized by
the $\phi$-deformed $E_{8}$ probe theory with maximal monodromy. An ``X'' indicates that
this entry has no meaning for the specified deformation.\label{table2}}
\end{table}
In comparison with the case of nilpotent deformations, we see that the central
charges and scaling dimensions shift very little. In the case $n = 9$, we find that a first
application of $a$-maximization yields a dimension for $Z$ below the unitarity bound. Applying
the prescription in \cite{Kutasov:2003iy}, we assume that this field decouples in the IR when
its dimension saturates the unitarity bound. Another curious feature of the above examples is that in the
case $n=3$, the cubic anomaly $a_\IR$ is quadratic in $t$ rather than cubic.
This means that the values of $t_{\ast}$ in this case will be rational
numbers, and all operator dimensions will also be rational. It would be
interesting to see whether there are any additional properties associated with
this behavior.

\subsection{$\mathbb{Z}_{2} \times \mathbb{Z}_{2}$ Monodromy}\label{E8z2z2}

As a simple case of potential phenomenological relevance, we now consider an
unfolding of $E_8$ down to $SU(5)_{\GUT} \times SU(5)_\perp$. Specifically, we consider
 a Dirac neutrino scenario of \cite{EPOINT}, where
$\phi \in SU(5)_\bot$ exhibits $\mathbb{Z}_{2} \times \mathbb{Z}_{2}$
monodromy. In
this case, the monodromy group acts by interchanging two pairs of eigenvalues
for $\phi$ independently. To illustrate the main point, we consider a
configuration with eigenvalues:%
\begin{equation}
\text{Eigenvalues}(\phi)=\left\{  a+\sqrt{z_{1}},a-\sqrt{z_{1}},b+\sqrt{z_{2}%
},b-\sqrt{z_{2}},-2a-2b\right\}
\end{equation}
where $a$ and $b$ are generic linear expressions in the $z_{i}$. A matrix
representative composed of two $2\times2$ blocks and one $1\times1$ block is:
\begin{equation}
\phi_\UV =
\begin{pmatrix}
& 1 \\
Z_1 - a^2 & 2a
\end{pmatrix}
\oplus
\begin{pmatrix}
& 1 \\
Z_2 - b^2 & 2b
\end{pmatrix}
\oplus
(-2a-2b).
\end{equation}
In the infrared, the deformation is characterized by:%
\begin{equation}
\phi_\IR=
\begin{pmatrix}
& 1 \\
Z_1 &
\end{pmatrix}
\oplus
\begin{pmatrix}
& 1 \\
Z_2 &
\end{pmatrix}
\oplus
(0).
\end{equation}
Note that in the infrared, the non-abelian flavor symmetry is $SO(10)$ rather
than $SU(5)$. Further, the characteristic equation for $\phi_\IR$ is:%
\begin{equation}
\left(  \phi_\IR^{2} - Z_{1}\right)  \left(  \phi_\IR^{2} - Z_{2}\right)
\phi_\IR=0.
\end{equation}

As for the case of maximal monodromy, the UV\ inputs are quite similar to the
case of the nilpotent deformations, though we also need to specify the values
of the parameters $\mu_{1}$ and $\mu_{2}$. In the case of the two Jordan
blocks, we have $r = 2$ and:
\begin{equation}
\mu_{1}  = \mu_{2} = \frac{5}{2}
\end{equation}
The IR values of the various central charges and dimensions are shown in Table~\ref{z2z2table}.
\begin{table}\[
\begin{array}
[c]{|c|c|c|c|c|c|c|c|c|c|c|c|}\hline
& t_{\ast} & a_\IR & c_\IR & k_\IR &
\left[  Z\right]   & \left[  Z_{1}\right]   & \left[  Z_{2}\right]   &
[\cO _\GUT ] & \left[  \cO _{i\overline{j}}^{\min}\right]   & [\cO _{i}^{\min}] &
[\cO _{ij}^{\min}]\\\hline\hline
E_{8} \,\,\rm{ with }\,\, \mathbb{Z}_{2} \times \mathbb{Z}_{2} & 0.46 & 3.11 & 3.95 & 8.32 & 4.16 & 1.39 &
1.39 & 2.31 & 1.61 & 1.96 & 1.61\\\hline
E_{6} \,\,\rm{ with }\,\, \mathbb{Z}_{2} & 0.51 & 1.45 & 1.80 & 4.58 & 2.29 & 1.53 & 1 & 2.24 & 1.47 & 1.85 & \XX  \\
\hline
\end{array}
\]

\caption{Central charges and dimensions of the cases with monodromy. The first row is for the deformation of the $E_8$ theory with $\bZ_2\times\bZ_2$ monodromy, discussed in section \ref{E8z2z2}.  The second row is for the $E_6$ theory with $\bZ_2$ monodromy, discussed in section \ref{E6z2}. \label{z2z2table}}
\end{table}

\subsection{$E_{6}$ and $\mathbb{Z}_{2}$ Monodromy}\label{E6z2}

The minimal requirement for a large top quark
Yukawa coupling is an $E_6$ point. Additionally, for one heavy generation, we require the unfolding to $SU(5)_\GUT$ to have $\bZ_2$ monodromy \cite{Hayashi:2009ge}.
To explicitly see the effects of the monodromy group, we consider the breaking
pattern:
\begin{align}
E_{6}  &  \supset SU(5)\times U(1)\times SU(2)\\
78  &  \rightarrow(\rep{24}_{0},\rep{1})+(\rep5_{6},\rep1)+(\overline{\rep5}_{-6},\rep1)+(\rep1_{0},\rep3)+(\rep{10}_{-3},\rep2)+(\overline{\rep{10}}_{+3},\rep2).
\end{align}
The monodromy group $\mathbb{Z}_{2}$ is the
Weyl group of $SU(2)$, which acts by interchanging the two
components of an $SU(2)$ doublet. A matrix $\phi$ taking values in
$SU(2)\times U(1)$ which accomplishes this is:%
\begin{equation}
\phi=\left(
\begin{array}
[c]{cc}
& 1\\
Z_{1} &
\end{array}
\right)  \oplus\left(  Z_{2}\right)  .
\end{equation}
In particular, using the general result of \cite{Green:2010da}, we see that the
$z_{2}$-dependent contribution drops out and only the $2\times2$ block of
$SU(2)$ dictates the flow in the IR.\ In the IR, the non-abelian flavor symmetry of the
CFT is $SU(6)$, and the dimension of $Z_{2}$ remains one. The value of the
parameter $\mu_{1}$ is:%
\begin{equation}
\mu_{1} = \frac{5}{2}.
\end{equation}
Computing the IR values of the various central charges and operator dimensions, we find the values shown in Table~\ref{z2z2table}.

\section{Stabilizing the D3-Brane Position}\label{sec:POLY}

In much of this paper we have focused on the effects of the deformation:%
\begin{equation}
\delta W=\Tr_{G}(\phi(Z_{1},Z_{2})\cdot \cO ). \label{phiO}%
\end{equation}
From the perspective of the field theory, additional deformations can be
built purely from $Z$, $Z_1$ and $Z_2$. In the context of a string compactification,
such superpotentials serve to stabilize the position of the D3-brane. These
position-dependent superpotential terms can be expanded in a power series in
the $Z_{i}$'s:%
\begin{equation}
W_{\text{position}}\left(  Z_{1},Z_{2},Z\right)  =F_{i}Z_{i}+M_{ij}Z_{i}%
Z_{j}+\lambda_{ijk}Z_{i}Z_{j}Z_{k}+\cdots \label{Wpos}%
\end{equation}
where we have set $Z_{3}=Z$, and the $F$, $M$ and $\lambda$ correspond to
constants of the theory, which in a string compactification would be moduli-dependent parameters. Such terms are expected to be generated
in the presence of appropriate fluxes, as studied for example in \cite{Martucci}
and \cite{FGUTSNC}.

This is already quite interesting for the purposes of model
building because superficially the term (\ref{Wpos}) suggests the superpotential of an
O'Raifeartaigh model. Caution is warranted, however, because the
fields $Z_i$ do not have a canonical K\"ahler potential. It is nevertheless tempting to speculate that the D3-brane could
naturally provide a source of supersymmetry breaking. Further, the fact that
there are operators with Standard Model gauge quantum numbers also suggests
that such a sector could naturally communicate supersymmetry breaking to the
visible sector via gauge mediation effects. A full analysis of these
possibilities is beyond the scope of the present work, and so in the remainder
of this section we confine our analysis to the question of which deformations lead
to another interacting CFT, when combined with the deformation of equation (\ref{phiO}).

In principle, we can consider either continuous or discrete symmetries which
exclude all such contributions, thus retaining the original form of the
CFT\ induced by just the deformation $\Tr_{G}(\phi\cdot \cO )$.
For an appropriate discrete subgroup of the remaining flavor symmetries
of the system, it is immediate that we can  exclude such linear
and quadratic deformations from appearing in equation (\ref{Wpos}). On the
other hand, for appropriate choices of a discrete symmetry such as $\mathbb{Z}_{2}$,
we can also consider cases where for example $Z_{1}$ is excluded, but $Z_{1}^{2}$ is allowed.

\paragraph{Linear term in $Z_1$ and $Z_2$}
Let us now suppose that at least some of the terms of $W_{\text{position}}$
are not forbidden by a discrete symmetry. We can see that some of these terms
could be relevant deformations because the deformation $\Tr_{G}(\phi(Z_{1}%
,Z_{2})\cdot \cO )$ increases the dimension of $Z_{1}$ and $Z_{2}$, but decreases
the dimension of $Z$.
Note that deformations linear in $Z_{1}$ and $Z_{2}$ do not
induce a flow to a CFT. The reason is that if we demand such operators are
marginal in the IR, then the operator $\cO _{(i)}$ multiplying $Z_{i}$ in the
superpotential deformation would have dimension zero. Indeed, in the UV theory it is clear that adding this term simply
enforces the condition that the vev of $\cO _{(i)}$ is non-zero. This is also
what is expected based on the brane construction. Viewing the probe D3-brane
as an instanton, the linear terms in the $Z_{i}$  fix some
of its moduli.

\paragraph{Linear term in $Z$}
Let us next consider  a term linear
in $Z$ in the $D_{4}$ probe theory. With notation as in Section 3, this corresponds to a mass term for the adjoint scalar
$\varphi$ which tends to attract the D3-brane to the seven-brane.
In the absence of the $\Tr_{SO(8)}(\phi\cdot \cO )$ deformation,
deforming by $Z$ is the standard adjoint mass deformation.
In the IR, the field $\varphi$ has been integrated out, leading to
an infrared marginal quartic interaction between the quarks. In the presence
of the further deformations by $\Tr_{SO(8)}(\phi\cdot \cO )$, we cannot
simultaneously demand that both $Z$ and these deformations are marginal in the
IR. To see this, consider again the case of constant nilpotent $\phi$ studied in section
\ref{AmaxNilp}. Integrating out the massive quarks requires that the operator
$Q_{1}\varphi^{n+1}\widetilde{Q}_{\overline{n}}$ be marginal in the IR. Note
that this is incompatible with the condition that $Z$ is also marginal. Thus,
starting from the original $\cN=2$ theory, we can either deform by
$\Tr_{SO(8)}(\phi\cdot \cO )$ or by $Z$, leading to two different interacting CFTs.
Adding both terms simultaneously, we see also that we cannot simultaneously
enforce that both deformations are marginal in the IR.

Next consider adding a linear term in $Z$ to the $\phi$-deformed E-type theories.
In the original $\cN=2$ $E_{6}$ theory, $Z$ is dimension three but is irrelevant in the
IR \cite{Green:2010da}. In the $E_{7}$ and $E_{8}$ theories, $Z$ has dimension $4$ and $6$ respectively,
and so is also irrelevant. In the $\phi$-deformed theories, however, the dimension of $Z$ can be significantly
lower. This in turn means that in some circumstances it is indeed appropriate to treat it as a
potentially marginal operator in the IR. For simplicity, let us consider the case of deformations by a nilpotent mass
deformation taking values in an $SU(n)$ subblock, and consider the further
effect of deforming the superpotential by a term linear in $Z$.

We now study whether such a deformation induces a flow to a new CFT. To this end, we assume
that the deformation linear in $Z$ is marginal in the IR, and deduce whether this leads to a consistent
picture of possible flows. In the absence of the term linear in $Z$, the IR R-symmetry was given in \eqref{riris}.
Requiring that $Z$ is also marginal in the IR determines $t$ via:%
\begin{equation}
t=\frac{2}{\Delta_\UV (Z)}.
\end{equation}
Let us demand that there is no violation of the unitarity bound for operators $\cO _{i\overline{j}}$.
The $T_{3}$ charge of the lowest spin component is $n-1$, which in turn means:%
\begin{equation}
R_\IR(\cO _{i\overline{j}}^{\min})=2-\frac{2}{\Delta_\UV (Z)}-\frac{2}%
{\Delta_\UV (Z)}\left(  n-1\right)  .
\end{equation}
Requiring $R_\IR(\cO _{i\overline{j}}^{\min})>2/3$ implies:
\begin{equation}
n<\frac{2\Delta_\UV (Z)}{3}.\label{baz}
\end{equation}
Note in particular that for the $D_{4}$ and $E_{6}$ theories, $\Delta
_\UV (Z)=2$ and $3$ respectively, and so this condition is not satisfied. For
the $E_{7}$ theory $\Delta_\UV (Z)=4$ and we can deform by an $SU(2)$
nilpotent subblock, while for the $E_{8}$ theory $\Delta_\UV (Z)=6$ and we can
deform by an $SU(3)$ or smaller subblock. In these cases, however, we observe that
the dimension of $Z$ in the original $\phi$-deformed theory is above three, and so the corresponding
deformation is irrelevant.

When \eqref{baz} is not satisfied, it is not clear whether to expect an
interacting CFT in the IR; the endpoint might instead be a massive theory.
It would be interesting to classify the available IR\ phases from the combined deformations induced
by $\Tr_{G}(\phi\cdot \cO )$ and $W_{\text{position}}$.

\section{Coupling to the Visible Sector}\label{sec:SM}

In much of this paper we have focused on the dynamics of the CFT sector,
providing evidence that close to the visible sector, there could be an
interacting $\cN = 1$ CFT. A full analysis of the various
consequences for phenomenology is beyond the scope of
the present paper, and so in this section we shall only make some general comments.

\subsection{Unfolding $E_{8}$ to $SU(5)_\GUT $} \label{ssec:Unfolder}

In this subsection we discuss in more
practical terms the deformations of an $E_{8}$ singularity down to an $SU(5)_\GUT $ singularity.
In practice, extracting the explicit form of an unfolding based on the local
form of $\phi$ can be somewhat cumbersome, in part
because the expressions for the primitive Casimir invariants of the E-type algebras are quite unwieldy
(see \cite{KatzMorrison} for their explicit forms). In the special case where $\phi$ takes values in the $SU(5)_{\bot}$
factor of the subalgebra $SU(5)_\GUT  \times SU(5)_{\bot} \subset E_{8}$, a significant simplification in the form
of the unfolding occurs.

In this case, $\phi$ has a characteristic equation of the form:
\begin{equation}
b_{0} \phi^{5} + b_{2} \phi^{3} + b_{3} \phi^{2} + b_{4} \phi + b_{5} = 0
\end{equation}
where the $b_{i}$ are holomorphic $z_{i}$-dependent coefficients. As
explained in \cite{Tatar:2009jk, DWIII} (see also \cite{BershadskyKachruSadov}), a local
unfolding of $E_8$ down to $SU(5)_\GUT $ can then be written as:
\begin{equation}
y^{2} = x^{3} + b_{0} z^{5} + b_{2} x z^{3} + b_{3} y z^{2} + b_{4} x^{2} z + b_{5} x y.
\end{equation}
From this family of curves we can read off the value of $\tau$ on the Coulomb branch,
and also the relative scaling
dimensions of mass deformations to the dimension of the Coulomb branch parameters.
We emphasize, however, that knowing the coefficients $b_{i}$ is \textit{not} enough to reconstruct a unique choice of $\phi$.

\subsection{Coupling of the CFT to the Visible Sector}

Now, note that the full system described by the CFT and the visible sector will no longer be a
CFT. Indeed, upon compactifying to four dimensions, the flavor symmetry will
be weakly gauged, and conformality will be lost. The matter fields of the
visible sector  can either localize on matter curves of the compactification,
or propagate in the bulk worldvolume of the seven-brane. Thus, we can in
principle study the effects of first compactifying the matter curves, and then
consider the additional effects of compactifying the remaining directions of
the seven-brane.

Matter fields $\psi_{R}$ transforming in a representation $R$ of $SU(5)_\GUT $ couple to
operators $\cO _{R^{\ast}}$ of the CFT transforming in the dual representation via:
\begin{equation}\label{coupling}
\int d^{2}\theta\text{ }\psi^{(i)}_{R}\cdot f_{(i)}(Z_{\parallel})\cO _{R^{\ast}}
\end{equation}
where $i = 1,...,3$ is a generational index for chiral matter, and $f_{(i)}(Z_{\parallel})$ is a
function of the local coordinate $Z_{\parallel}$ for the matter curve. Though
the specific details of the couplings depend on seven-brane monodromy,
the main point is that this adds another class of interactions to consider,
which it would be interesting to analyze further. Since the
matter field wave functions have different profiles near a Yukawa point,
the order of vanishing near this point will dictate the relevance of the
coupling to the visible sector. For example the coupling to the third (heavy) generation quarks
will be most relevant, while the coupling to the first generation will be least relevant.

The CFT also possesses a large number of states which are charged under the
GUT group $SU(5)_\GUT $. These states will in turn affect the running.
As explained in \cite{FunParticles}, the contribution to
the running is essentially fixed by the scaling of the Coulomb branch
parameter $\Delta_\IR(z)$.  \def\ff{\rep5\oplus\overline{\rep5}}
More directly,  the contribution to the one-loop running effects of the $SU(5)$ GUT coupling  from the CFT is the same as $N_{\ff}$ pairs of $\ff$ chiral multiplets,
where
\begin{equation}
N_{\ff}=\frac{k_\IR}{2}.
\end{equation}
Scanning over the values of $k_\IR$ we have already computed, we see that in
the case of the $\cN=2$ $E_{8}$ theory, this has the effect of six
$\ff$'s. On the other hand, in the case of larger deformations
down to $SU(5)$, we see what would appear as an irrational
number of $\ff$'s, with the net effect on the order of two
$\ff$'s in the case of maximal monodromy.

The study of how this sector couples to the visible sector likely has a rich
phenomenology which could be studied further for various model
building applications.

\section{Conclusions}\label{sec:CONC}

Recent work has shown that compactifications of F-theory provide a natural arena for engineering
gauge theories of potential phenomenological interest. In this paper we have found a new class of SCFTs
which arise as the worldvolume theories of D3-branes probing F-theory seven-branes, which
in appropriate circumstances can couple to the visible sector of the Standard Model. These SCFTs are
characterized in terms of deformations of an $\cN = 2$ system. In many cases, we have argued that the
resulting deformation induces a flow to a new interacting SCFT. We have also seen that while the geometry of the seven-branes of F-theory is able to capture a great deal of information about such theories, in particular through the $\cN = 1$ curve,
additional input from $a$-maximization is often necessary to fully specify the infrared R-symmetry. These CFTs are also of potential phenomenological relevance, as the states of the D3-brane theory can couple to the Standard Model. In the rest of this section we discuss some future avenues of
potential investigation.

One of the central themes of this work has been the role of backgrounds in which $[\phi, \phi^{\dag}] \neq 0$.
This suggests a sense in which the seven-branes of F-theory could ``puff up'' to non-commutative nine-branes. Non-commutativity in
F-theory compactifications has recently been discussed in \cite{FGUTSNC,HeckVerlinde}. It would be
worthwhile to develop a more uniform treatment of F-theory from the non-commutative viewpoint.

In much of this work, we have only been able to provide various consistency checks that the $\cN = 1$ theories
flow to an interacting SCFT. It would be interesting to develop further consistency checks of these statements. Along these lines,
it would be useful to develop a holographic dual description of these SCFTs in the large $N$ limit. In the $\cN = 2$
setting, holographic duals are available which have been studied for example in \cite{Aharony:1998xz}. In addition, we have
argued that further deformations can restore the system to an $\cN = 2$ system
which also admits a holographic dual. It would be quite instructive to study whether there is an interpolating $\cN = 1$
geometry which connects these $\cN = 2$ theories.

Another potential avenue would be to search for possible field theory duals of the theories considered
here. Indeed, some notable examples for related $\cN=2$ theories have been studied for example in \cite{Argyres:2007cn, Argyres:2007tq}, and it would
be interesting to see whether $\cN = 1$ analogues of these duals could be constructed.

Aside from providing further consistency checks, it would also be enlightening to further study the structure of these new SCFTs.
For example, determining the chiral ring for these theories, or even the number of independent generators for the chiral
ring (perhaps along the lines of \cite{Benvenuti:2010pq}) would be quite helpful. Determining an index similar
to the one recently computed in \cite{Gadde:2010te} for related $\cN = 2$ theories would also be of interest.

Finally, though our main focus in this paper has been the study of the associated SCFTs,
we have also seen that some of the main ingredients present in the D3-brane probe theory
could potentially be used for breaking supersymmetry. Further, since the
CFT comes equipped with fields charged under the visible sector gauge group, it is quite
natural to speculate that the D3-brane already contains all the ingredients to realize
a self-contained gauge mediation sector.

\section*{Acknowledgements}

We thank C. C\'{o}rdova, D. Green, K. Intriligator, Y. Ookuchi, S-J. Rey, and N. Seiberg
for helpful discussions. We thank the 2010 Simons workshop in Mathematics and Physics
for hospitality during part of this work. JJH also thanks the Harvard high energy
theory group for hospitality during part of this work. The work of JJH and YT
is supported by NSF grant PHY-0503584. The work of CV is supported by
NSF grant PHY-0244821. The work of BW is supported by DOE grant DE-FG02-90ER40542. YT is additionally supported by the Marvin L. Goldberger membership at the Institute for Advanced Study. BW is additionally supported by the Frank and Peggy Taplin Membership at the
Institute for Advanced Study.

\appendix
\section{Field Theory Tools}\label{subsec:FTT}
Here we collect standard facts on four-dimensional SCFTs. Detailed discussions can be found in references.

First, there is a lower bound to the dimensions $\Delta$ of the operators of a unitary CFT.
For example, all spin zero gauge-invariant operators which are not free fields must satisfy $\Delta > 1$.
For scalar chiral primaries of a SCFT,
the dimension of an operator is related to its R-charge $R$ through the relation
\begin{equation}
\Delta = \frac{3}{2} R.
\end{equation}
To extract the scaling dimensions of operators,  it is therefore of interest to
determine the infrared R-charge $R_\IR$ of the operators. Note that finding
$R_\IR<2/3$ would imply that we had made an incorrect assumption. Ensuring all operators are
above the unitarity bound therefore provides a basic check on our analysis.

In the context of $\cN = 2$ theories, it is often possible to fix $R_\IR$ by using the $\cN = 2$ Seiberg-Witten curve, see \cite{Argyres:1995jj,Argyres:1995xn}. For example, using this procedure it is possible to fix the scaling of the Coulomb branch parameter
for the $\mathcal{N} = 2$ SCFTs realized by a D3-brane probing an F-theory singularity at constant axio-dilaton.
For $\cN = 1$ theories the situation is more complicated because we no longer
have the analogue of the  $\cN = 2$  BPS bound.\footnote{In
\cite{Aharony:1996bi} it was suggested that the special geometry of the Calabi-Yau of F-theory
could be used to fix the scaling dimensions of operators. As can be shown, this is equivalent to demanding
the Gukov-Vafa-Witten flux induced superpotential of \cite{GukovVafaWitten} has dimension exactly three. However, this
superpotential also significantly alters the theory, rendering this method quite suspect.}
To fix the scaling dimensions we use $a$-maximization \cite{Intriligator:2003jj}.

Let $R_0$ denote an R-symmetry, i.e. a symmetry under which the supercharge has
charge 1. Let $F_I$ be the generators of the abelian flavor symmetries.
Then $R_\IR$, the R-symmetry of the IR theory in the superconformal algebra can be written as
\begin{equation}
\label{rir}
R_\IR = R_{0} + \sum t_{I} F_{I}.
\end{equation}
The central charges $a$ and $c$ are then given in terms of 't Hooft anomalies by
\begin{align}
a_\IR &= \frac{3}{32}[3 \Tr R_\IR^3 - \Tr R_\IR],  &
c_\IR &= \frac{1}{32}[9 \Tr R_\IR^{3} - 5 \Tr R_\IR]. \label{boo}
\end{align}
The $a$-maximization procedure states that the $t_I$ can be determined
by first promoting $a_\IR$ to a function of $t_I$ by putting \eqref{rir} into \eqref{boo},
and then finding the unique local maximum of $a_\IR(t_{I})$.

A practical difficulty which is always encountered is to list all the infrared
flavor symmetries $F_I$, which might include emergent symmetries in the IR.
Assuming there is no such emergent symmetry, the procedure is to find a candidate $R_0$
by requiring the vanishing of the gauge anomaly, and then to demand
that any operators used to deform the superpotential become marginal in the IR and thus have  $R_\IR=2$,
with which we can extract relations between the parameters $t_{I}$.
The remaining parameters are then determined by $a$-maximization.

When the resulting maximum leads to one or more gauge-invariant operators  of
dimension less than one, this means our original assumption is wrong.
One interpretation  is that an emergent $U(1)$ appears in
the IR which acts only on the operator which seems to violate the unitarity bound,
making it  saturate the unitarity bound and decouple as a free field instead
\cite{Kutasov:2003iy} (see also \cite{Intriligator:2003mi}). The procedure
is to perform $a$-maximization again, but with the contribution from this operator removed:
\begin{equation}
a_{\text{new}}=a_{\text{old}}-\frac{3}{32}\left[  3\left(  r-1\right)  ^{3}-\left(
r-1\right)  ^{3}\right]  +\frac{1}{48}.
\end{equation}
where $r$ is the R-charge of the operator computed with respect to the old
R-charge assignments.

We can also check if additional deformations of the CFT decrease the value of $a_\IR$.
Physically, this is a reasonable condition to hope for, as the central charges can
be viewed as roughly counting the number of degrees of freedom of the CFT.

In addition to the central charges $a$ and $c$, there are central
charges associated with  flavor symmetries. Given flavor symmetry currents
$J_{A}$ and $J_{B}$, with $A,B$ indices labelling the generators of the
non-abelian flavor symmetry, the cubic anomaly
\begin{equation}
\Tr(R_\IR J_{A}J_{B})=-\frac{k_\IR}{6}\delta_{AB}\label{k}
\end{equation}
determines the effect of the CFT on the running of the holomorphic gauge
coupling of a weakly gauged flavor symmetry group.

In any $\cN=2$ conformal theory, there is an R-symmetry $SU(2)\times U(1).$ Denote by $I_{3}$ the
Cartan generator of the $SU(2)$ factor, and $R_{\cN=2}$ the generator
of the abelian factor. One linear combination of these generators is the $\cN=1$
R-symmetry $R_{\cN=1}$ and another corresponds to
a flavor symmetry $J_{\cN=2}$ as an $\cN=1$ SCFT:%
\begin{align}
R_{\cN=1}  &  =\frac{1}{3}R_{\cN=2}+\frac{4}{3}I_{3}, &
J_{\cN=2}  &  =R_{\cN=2}-2I_{3}.
\end{align}
$\cN=2$ supersymmetry relates the anomalies with the central charges as follows:
\begin{align}
\Tr(R^{3}_{\cN=2})  &  =\Tr(R_{\cN=2})=48(a_\UV -c_\UV ) ,\label{AAA}\\
\Tr(R_{\cN=2}I_{3}I_{3})  &  =4a_\UV -2c_\UV,\label{BBB}\\
\Tr(R_{\cN=2}J^{A}J^{B})  &  =-\frac{k_\UV }{2}\delta^{AB}.\label{CCC}
\end{align}

The central charges for the $\mathcal{N} = 2$ SCFTs realized by $N$ D3-branes probing an
F-theory singularity with constant axio-dilaton have been determined
\cite{Cheung:1997id,Aharony:2007dj}:
\begin{align}
a  &  = \frac{1}{4} N^{2} \Delta + \frac{1}{2} N (\Delta - 1) - \frac{1}{24},\\
c  &  = \frac{1}{4} N^{2} \Delta + \frac{3}{4} N (\Delta - 1) - \frac{1}{12},\\
k  &  = 2 N \Delta.
\end{align}
where $\Delta$ is the dimension of the Coulomb branch parameter $Z$ (see table \ref{tableTIME}). Note
that in the above formulae for $a$ and $c$ the contribution from the hypermultiplet $Z_{1} \oplus Z_{2}$ has
been subtracted off.

\baselineskip=.95\baselineskip
\bibliographystyle{utphys}
\bibliography{fcft}

\end{document}